\documentclass[aps,pre,twocolumn,groupedaddress,11pt,tightenlines]{revtex4-1}
\usepackage{amsmath,amssymb}
\usepackage{graphicx}
\usepackage{color} 
\usepackage{textcomp} 
\usepackage{comment}
\usepackage{subcaption}
\usepackage{multirow}
\usepackage{flushend}
\usepackage[title,titletoc,toc]{appendix}
\usepackage[justification=RaggedRight]{caption}

\DeclareMathOperator*{\argmin}{arg\,min}
\begin{document}
\title{To react or not to react? Intrinsic stochasticity of\\ human control in virtual stick balancing}
 
\author{Arkady Zgonnikov}
\email[]{arkady.zgonnikov@gmail.com}

\author{Ihor Lubashevsky}
\email[]{i-lubash@u-aizu.ac.jp}
 
\author{Shigeru Kanemoto}
\author{Toru Miyazawa}
\author{Takashi Suzuki}

\affiliation{University of Aizu, Tsuruga, Ikki-machi, Aizuwakamatsu, Fukushima 965-8580, Japan} 

\begin{abstract}
Understanding how humans control unstable systems is central to many research problems, with applications ranging from quiet standing to aircraft landing. Increasingly much evidence appears in favor of event-driven control hypothesis: human operators only start actively controlling the system when the discrepancy between the current and desired system states becomes large enough. The event-driven models based on the concept of threshold can explain many features of the experimentally observed dynamics. However, much still remains unclear about the dynamics of human-controlled systems, which likely indicates that humans employ more intricate control mechanisms. The present paper argues that control activation in humans may be not threshold-driven, but instead intrinsically stochastic, noise-driven. Specifically, we suggest that control activation stems from stochastic interplay between the operator's need to keep the controlled system near the goal state on one hand and the tendency to postpone interrupting the system dynamics on the other hand. We propose a model capturing this interplay and show that it matches the experimental data on human balancing of virtual overdamped stick. Our results illuminate that the noise-driven activation mechanism plays a crucial role at least in the considered task, and, hypothetically, in a broad range of human-controlled processes.
\end{abstract}

\maketitle

\section{Introduction}
Control of unstable systems underlies many critical procedures performed by human operators (e.g., manipulation of industrial machinery, aircraft landing~\cite{endsley1995toward}), as well as numerous routines all of us face in daily life (e.g., standing upright~\cite{collins1993open}, riding a bicycle~\cite{jones1970stability}, carrying a cup of coffee~\cite{mayer2012walking}). Eliciting and modeling the basic mechanisms of human control can help us to understand the nature of such processes, and in the end, hopefully, to reduce the risks associated with human error~\cite{reason1990human, moss2003balancing}.

Continuous control models describe human actions well in many situations~\cite{van2007postural, gawthrop2009predictive, gawthrop2011intermittent}. On the other hand, increasingly much evidence appears in favor of a more general concept, intermittent control~\cite{gawthrop2011intermittent, loram2011human, balasubramaniam2013control, milton2013intermittent, asai2013learning}. As far as human behavior is concerned, intermittency implies discontinuous control, which repeatedly switches off and on instead of being always active throughout the process. Intermittency has long been attributed to a general class of human-controlled processes~\cite{craik1947theory}. Nonetheless, despite being recognized for decades, human control intermittency is still far from being completely understood.

One of the most promising approaches to human control is event-driven intermittency, which claims that the control is activated when the discrepancy between the goal and the actual system state exceeds certain threshold. Models based on the notion of threshold can explain many features of the experimentally observed dynamics~\cite{gawthrop2011intermittent, milton2013intermittent}. However, much still remains unclear even in case of relatively simple control tasks, such as real~\cite{cabrera2002onoff,milton2009balancing,balasubramaniam2013control} or virtual~\cite{foo2000functional, bormann2004visuomotor, loram2009visual, loram2011human} stick balancing. For instance, the generating mechanism behind extreme fluctuations of the systems under human control (resulting, e.g., in stick falls) still has to be explained~\cite{cabrera2012stick}. Supposedly, more advanced mathematical concepts capturing core mechanisms of human control can contribute to deeper understanding of anomalous properties of human-controlled systems.

In the present paper we develop a notion of \textit{noise-driven control activation} as a more advanced alternative to the conventional threshold-driven activation. We argue that the proposed mechanism plays a key role in the fluctuations of unstable systems under human control. In our investigations we appeal to a novel experimental paradigm: balancing an overdamped inverted pendulum. The overdamping eliminates the effects of inertia and therefore reduces the dimensionality of the system. Arguably, the fundamental properties and mechanisms of human control are more likely to clearly manifest themselves in such simplified setup rather than in more complicated conventional experimental paradigms. Based on the insights provided by the experimental results, we elaborate a model implementing noise-driven control activation. The model captures the stochastic interplay between the operator's need to keep the stick upright and the inclination to halt the control (e.g., due to energy considerations). We then demonstrate that the model reproduces well the experimentally observed behavior. Our results suggest that the noise-driven control activation mechanism may be one of the key factors behind complex dynamics of human-controlled processes. 

\section{Methods}
Ten right-handed healthy volunteers (six male, four female, median age 26) participated in the experiments. Three subjects (labeled $1$ to $3$ in what follows) had previously participated in the preliminary experiments involving the same task~\cite{zgonnikov2012computer,zgonnikov2013dynamical}. Seven other participants had had no prior experience in either virtual or real stick balancing. All subjects gave written informed consent to participate in the experiments. Experimental procedures were approved by the University of Aizu Ethics Committee. 

\begin{figure}
\begin{center}
\includegraphics[width=0.4 \columnwidth]{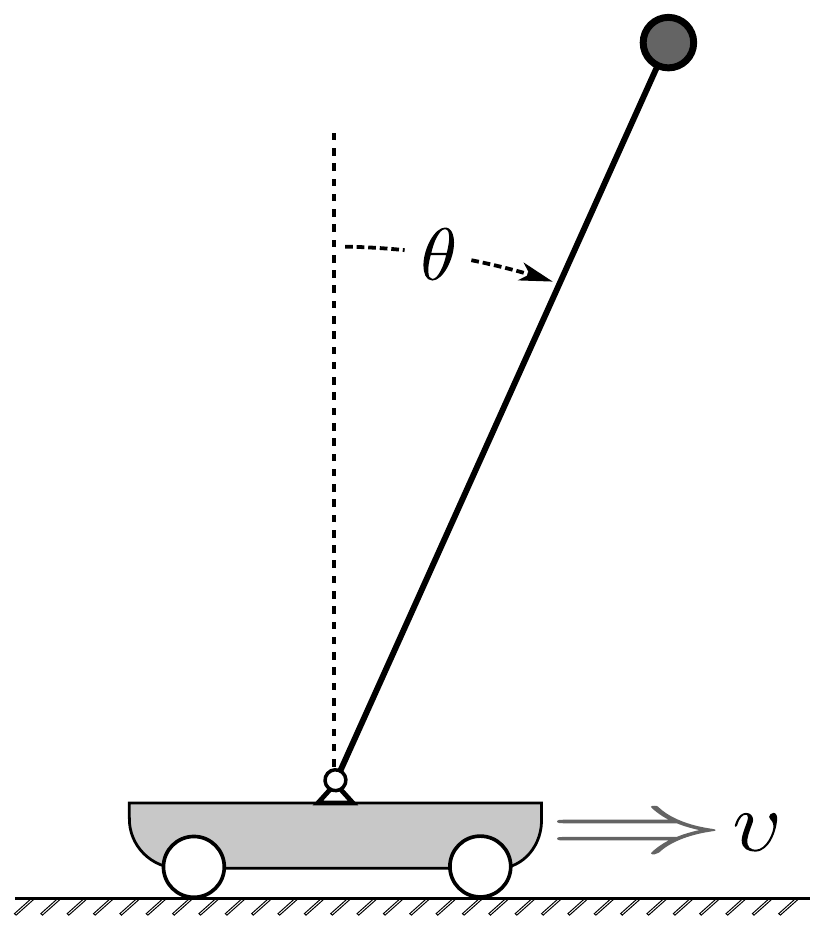}
\end{center}
\caption{
One-degree-of-freedom overdamped inverted pendulum. 
}
\label{fig:stick}
\end{figure}

The participants performed the task sitting at the office desk, using the common desktop computer. On the computer screen a subject saw a vertically oriented stick and a moving cart rigidly connected to the base of the stick (Fig.~\ref{fig:stick}). The task was to maintain the upright position of the stick by moving the platform horizontally via computer mouse. The data were collected in two experimental conditions corresponding to slow and fast motion of the stick (the slow stick task was offered first). For each condition the experiment consisted of one-minute practice period and three five-minute recorded trials separated by two three-minute rest periods. In the case of stick fall the initial system position was restored (platform put in the middle of the screen and the stick angle set to a small random value) and the subject was asked to click the button on the screen to continue the task. The distance between the monitor and the subject eyes was about 70 cm, the stick length on the screen was about 10 cm. The screen update frequency was 60 Hz. The horizontal position of mouse cursor on the screen was sampled with frequency of 50 Hz. A commercially available high-precision gaming mouse (Logitech G500) was used in the experiments.

The stick dynamics were simulated by numerically solving the ordinary differential equation (see Appendix~\ref{app:stick} for derivation)
\begin{equation}\label{eq:stick}
\tau \dot \theta = \sin \theta - \frac{\tau}{l} \upsilon \cos \theta\,,
\end{equation}    
where $\theta$ is the angular deviation of the stick from the vertical position and $\upsilon$ is the cart velocity. The parameter $\tau$ defines the time scale of the stick motion: the higher the $\tau$, the faster the stick falls in the absence of human control. The stick length $l$ de facto determines the characteristic magnitude of the cart displacements required for keeping the stick upright. The higher the $l$, the larger the cart velocity needed to compensate for certain stick deviation, and, consequently, the larger the typical amplitude of the cart motion. In the course of experiments the parameter $l$ modulated the relative impact of the mouse velocity on the stick dynamics, whereas the visible stick length on the screen was fixed.

The cart position was controlled by the operator via a computer mouse. Prior to each screen update the approximate horizontal mouse cursor velocity was calculated based on five most recent values of cursor position using the second-order low-noise differentiator~\cite{savitzky1964smoothing}. The resulting cursor velocity $\upsilon$ (measured in pixels per millisecond) was then substituted into Eq.~\eqref{eq:stick} which in turn was integrated using the first-order explicit Euler method~\cite{euler1768institutionum} to obtain the updated stick angle~$\theta$. 

Two combinations of stick parameters (see Table~\ref{tab:parameters}) were used in the experiments, representing the slow and fast stick dynamics. The fast stick parameters were tuned in such a way that the subjects had to remain steadily concentrated on the task in order to balance the stick successfully. On the other hand, the slow stick balancing was intended to be an easy, even boring task requiring few efforts from the operator. 
\begin{table}[!t]
\centering
\caption{
Fast and slow stick conditions}
\begin{tabular}{cccc}
Condition & $\tau$ & $l$\\
\hline
Slow & $0.7$ & $1.0$  \\
Fast & $0.3$ & $0.4$ \\
\end{tabular}
\label{tab:parameters}
\end{table}

To characterize the subjects in terms of their performance and balancing traits, three measures were used: 1) the average number of stick falls per minute $n_{\textrm{fall}}$; 2) the standard deviation of the stick angle $\textrm{std}(\theta)$ and 3) the proportion of total experimental time $\%_\textrm{drift}$ the mouse velocity $\upsilon$ was equal to zero. The first two measures, $n_{\textrm{fall}}$ and $\textrm{std}(\theta)$, reflect the subjects' balancing skill, whereas $\%_\textrm{drift}$ supposedly quantifies the intermittency of the subjects' control.

The model proposed in this study is represented by a set of stochastic differential equations. The numerical simulation of the model dynamics was performed using the explicit order 1.5 stochastic Runge-Kutta method~\cite{rossler2005explicit}. The simulation step $\Delta t = 0.01$ was chosen in such a way that varying it ten-fold could not affect the results of the simulation.

\section{Results}
\begin{table}
\centering
\caption{Balancing characteristics of the subjects. $n_{\textrm{fall}}$ is the average number of stick falls per minute, $\textrm{std}(\theta)$ is the standard deviation of the stick angle, and $\%_\textrm{drift}$ is the proportion of total balancing time the mouse velocity $\upsilon$ was equal to zero. In the slow stick condition no stick falls have been registered in all subjects. }

\begin{tabular}{ccc|ccc|cc}
\hline
\multirow{2}{*}{Subject} & \multirow{2}{*}{Sex} & \multirow{2}{*}{Age} & \multicolumn{3}{c}{Fast stick} & \multicolumn{2}{c}{Slow stick}                                       \\
\cline{4-6} \cline{7-8}
                         &                      &                      & $\textrm{std}(\theta)$ & $n_{\textrm{fall}}$ & $\%_{\textrm{drift}}$ & $\textrm{std}(\theta)$ &  $\%_{\textrm{drift}}$ \\
\hline
1                        & M                    & 22                   & 0.07                   & 0.00                & 42\%               & 0.03                   & 62\%               \\
2                        & M                    & 21                   & 0.21                   & 1.87                & 22\%               & 0.04                    & 48\%               \\
3                        & M                    & 25                   & 0.19                   & 0.93                & 25\%               & 0.07                    & 45\%               \\
4                        & F                    & 61                   & 0.36                   & 6.40                & 31\%               & 0.04                    & 59\%               \\
5                        & M                    & 20                   & 0.32                   & 3.67                & 10\%                & 0.12                   & 16\%               \\
6                        & M                    & 58                   & 0.38                   & 5.73                & 31\%               & 0.08                   & 46\%               \\
7                        & F                    & 27                   & 0.25                   & 2.73                & 35\%               & 0.03                    & 59\%               \\
8                        & M                    & 29                   & 0.18                   & 0.93                & 36\%               & 0.03                    & 56\%               \\
9                        & F                    & 58                   & 0.32                   & 4.93                & 31\%               & 0.04                   & 43\%               \\
10                       & F                    & 21                   & 0.28                   & 4.27                & 25\%               & 0.06                    & 37\%         \\    
\hline
\end{tabular}
\label{tab:subjects}
\end{table}

\begin{figure}[!ht]
\begin{center}
\includegraphics[width=1.0 \columnwidth]{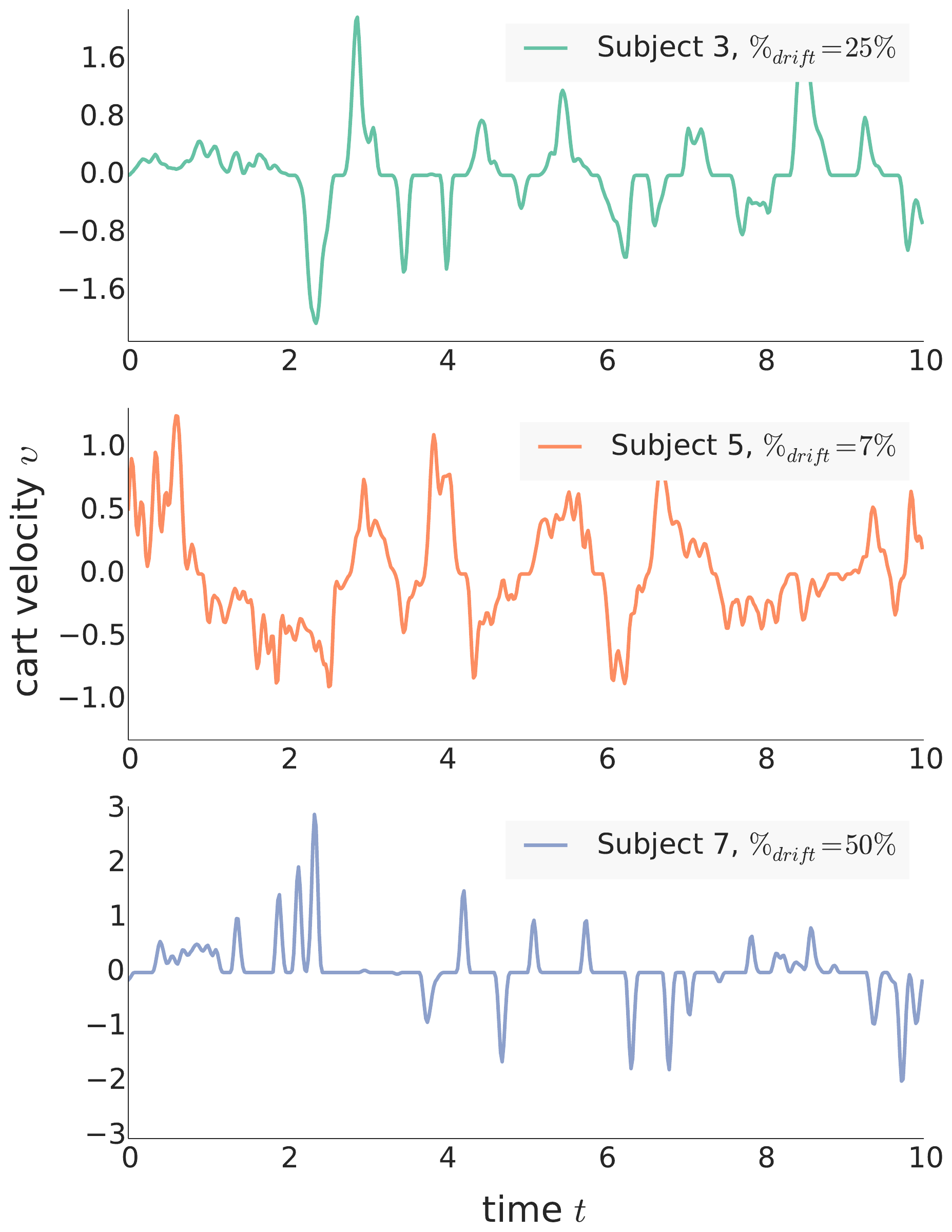}
\end{center}
\caption{
Cart velocity dynamics of three representative subjects. Each trajectory represents the randomly selected 10-second period of fast stick balancing without stick falls. The values of $\%_\textrm{drift}$ (calculated based on the presented 10-second time series) are shown for reference.
}
\label{fig:exp_velocity}
\end{figure}

\begin{figure}[!ht]
\begin{center}
\includegraphics[width=1.0 \columnwidth]{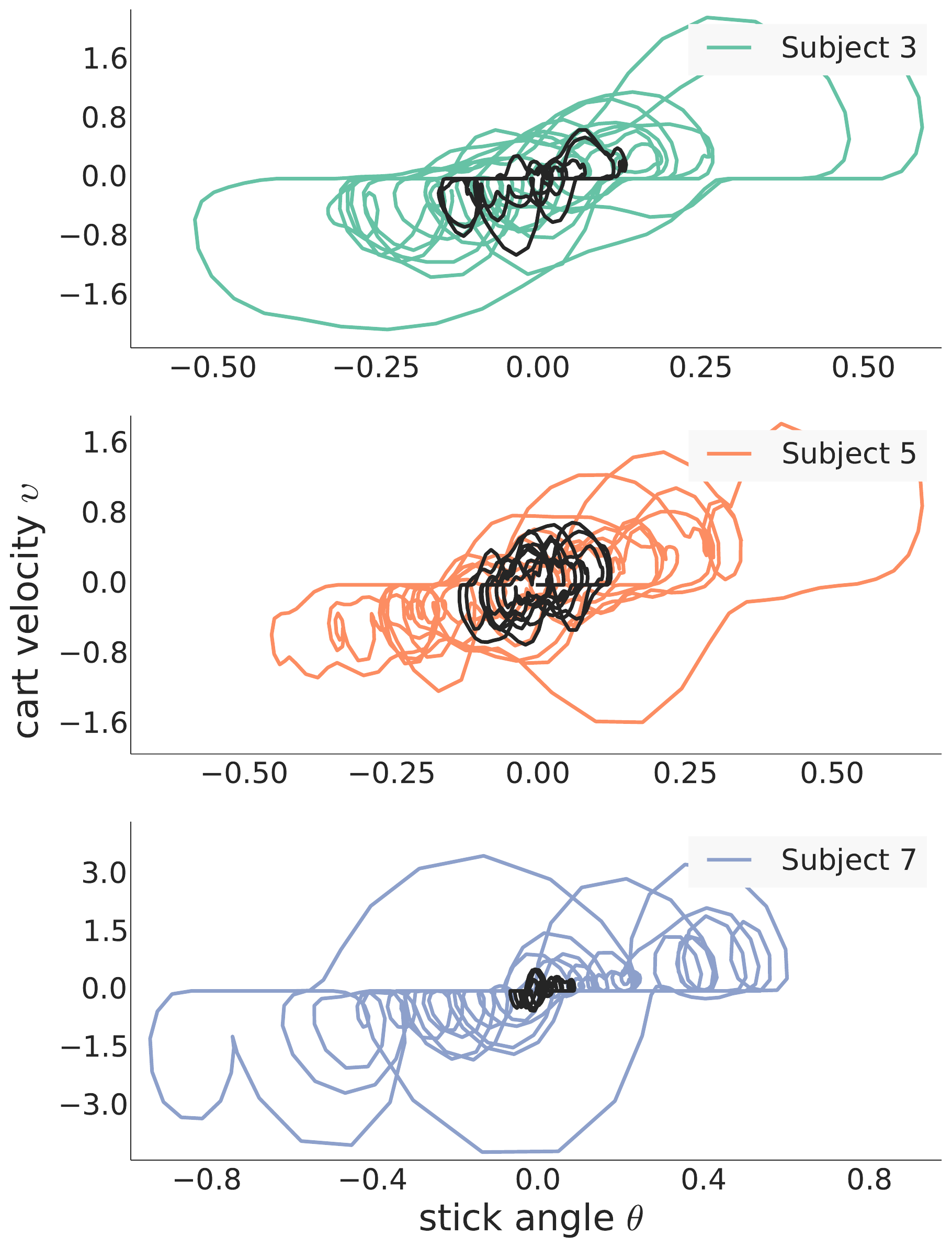}
\end{center}
\caption{
Phase trajectories of the overdamped stick balancing. Colored trajectories correspond to the fast stick condition and represent the randomly selected 15-second time fragments. Same subjects' trajectories obtained in the slow stick condition are shown in black.
}
\label{fig:exp_phase_three}
\end{figure}

The subjects' performance varied greatly across the two conditions (Table~\ref{tab:subjects}). In the slow stick condition no stick falls have been registered in all subjects, and $\textrm{std}(\theta)$ remained consistently small (median $0.04$). The fast stick condition revealed the diversity of the subjects with respect to their balancing skill: the least skilled of them (Subjects 4 and 6) could not balance the stick longer than 10 seconds on average, whereas the expert one (Subject 1) handled the task remarkably well. In the fast stick condition two skill indicators, $\textrm{std}(\theta)$ and $n_{\textrm{fall}}$, correlated significantly with each other ($r=0.948,\,p=0.00003$) and with the age of the subjects ($r=0.68,\,p=0.03$ for $\textrm{std}(\theta)$, $r=0.755,\,p=0.012$ for $n_{\textrm{fall}}$). One of the specific questions for the further analysis is whether or not the basic properties of the ``relaxed'' and ``effortful'' regimes of human control (corresponding to the slow and fast stick condition respectively) are different. In what follows we focus on the fast stick condition, mentioning the complementary results for the slow stick task where appropriate.

Pronounced intermittent control patterns were found in all but one subjects regardless of their skill. Average value of $\%_\textrm{drift}$ in the fast stick condition fell in the range of $22\%$ to $42\%$ for all participants except Subject~5. In the slow stick task $\%_\textrm{drift}$ was consistently greater compared to the fast stick, and correlated negatively with $\textrm{std}(\theta)$ ($r=-0.885,\,p=0.0006$). Interestingly, we did not find any relationship between $\%_\textrm{drift}$ in the fast stick condition and subjects' $\textrm{std}(\theta)$, $n_{\textrm{fall}}$, age or previous experience.

The observed intermittency is illustrated by the typical cart velocity dynamics (Fig.~\ref{fig:exp_velocity}). Subjects 3 and 7 control the stick intermittently: they spend substantial portion of time in the passive control phase. The active control fragments are often short, unimodal and isolated, which prompts that the subjects employed open-loop rather than feedback control. The control strategy exhibited by Subject~5 is seemingly of different, continuous nature. Although the multimodal active control fragments comprising several consecutive corrections are also present in other subjects, there is practically no passive periods in the velocity profile produced by Subject~5. Whether such a difference in the subjects' control strategies contributes considerably to the task dynamics is to be investigated below.

\begin{figure*}
\begin{center}
\includegraphics[width=1.0 \textwidth]{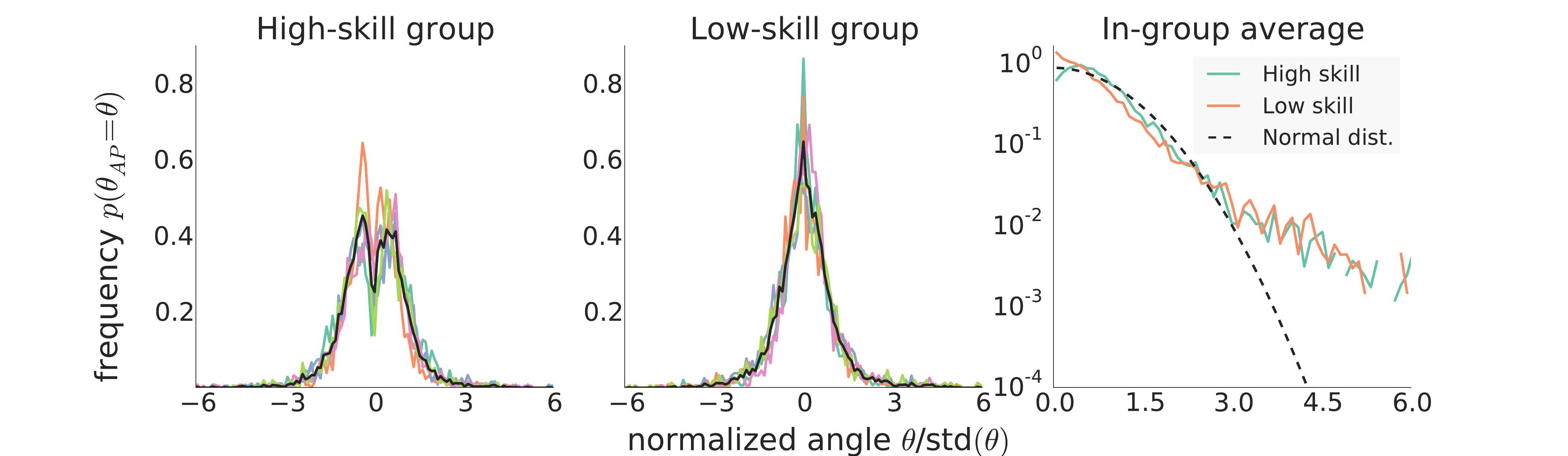}
\end{center}
\caption{
Distribution of the action points (the values of stick angle triggering the operators' response) in the fast stick condition. The angle value was counted as an action point if it corresponded to the instant the mouse velocity switched from zero to a non-zero value. In the left and middle frames colored lines represent the distributions for each subject. Black lines represent the average distributions across each group. The right frame illustrates the average distributions of the absolute value of action point for each group of subjects in the logarithmic scale; the standard normal distribution truncated at zero is represented for reference. The high skill group consists of Subjects 1, 2, 3, 7, 8; the low skill group includes Subjects 4, 5, 6, 9, 10.
}
\label{fig:exp_ap_distr}
\end{figure*}

The phase space of the standard, underdamped inverted pendulum includes two independent variables, $\theta$ and $\dot \theta$. In contrast, the dynamics of the overdamped stick in the absence of external forces can be completely described solely by the stick angle $\theta$. This allows us to graphically represent the dynamics of the task at hand by considering a hypothetical dynamical system comprising two independent yet coupled components: the overdamped stick and the human operator. The phase space of this system should then include, first, the stick angle $\theta$, and, second, the cart velocity $\upsilon$ as a separate phase variable characterizing the operator's actions. The trajectories of the stick balancing in the $\theta\upsilon$ phase plane provide important insights into the system dynamics (Fig.~\ref{fig:exp_phase_three}).

\begin{figure*}
\centering
\begin{subfigure}{0.47\textwidth}
\includegraphics[width=\textwidth]{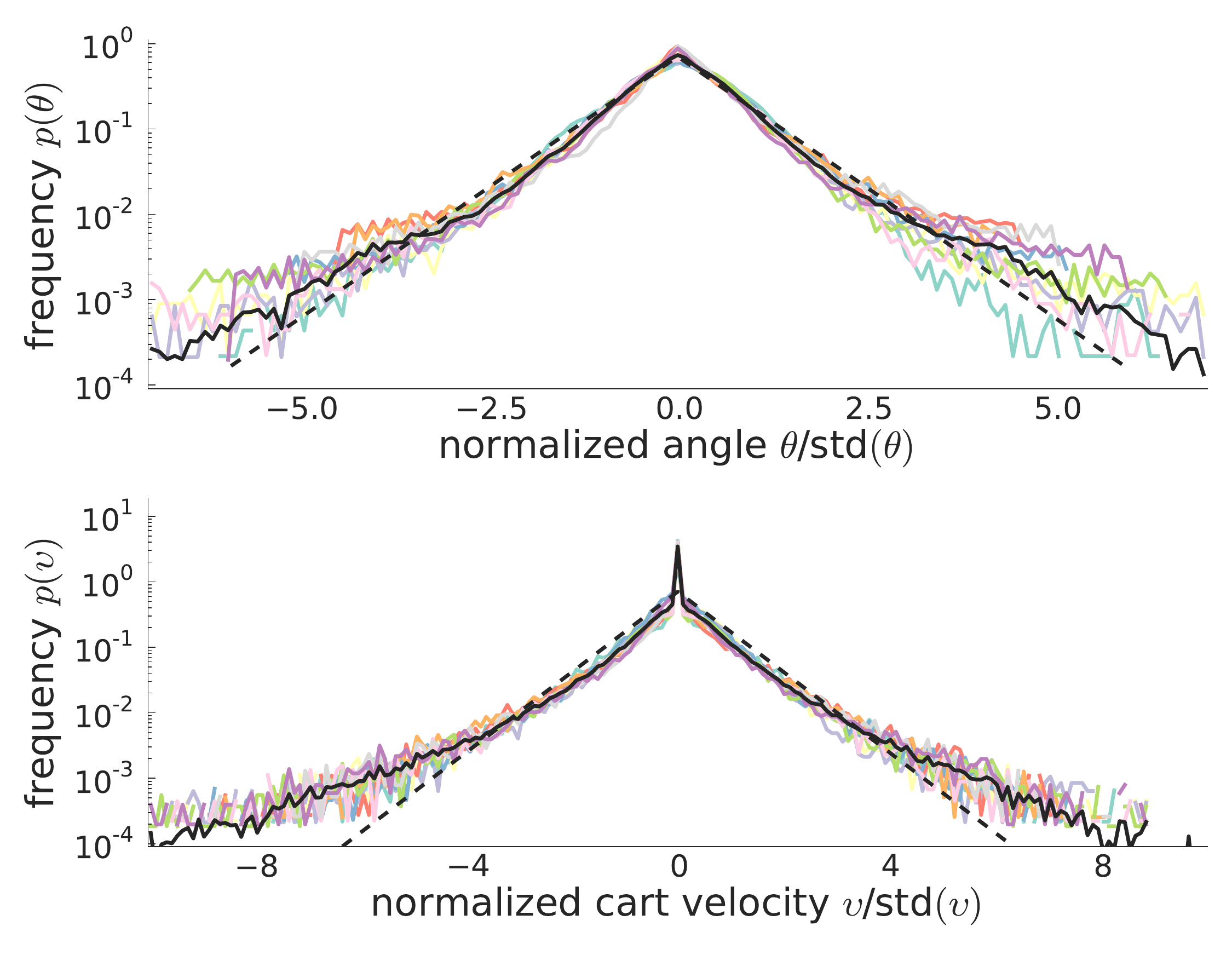} 
\caption{Fast stick: logarithmic scale}
\label{fig:exp_distr_fast_log}
\end{subfigure}
\quad
\begin{subfigure}{0.47\textwidth}
\includegraphics[width=\textwidth]{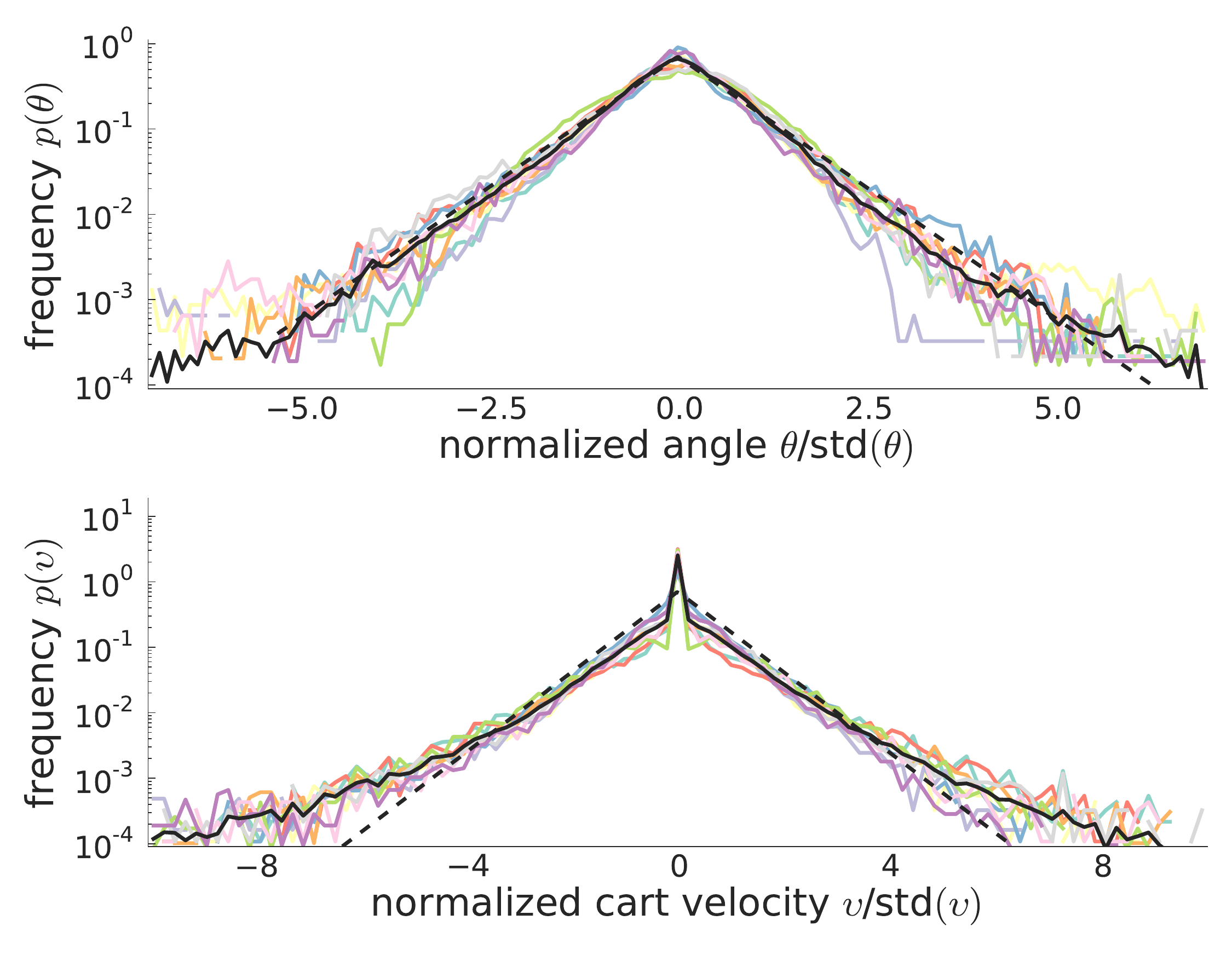} 
\caption{Slow stick: logarithmic scale}
\label{fig:exp_distr_slow_log}
\end{subfigure}
\\ 
\begin{subfigure}{0.47\textwidth}
\includegraphics[width=\textwidth]{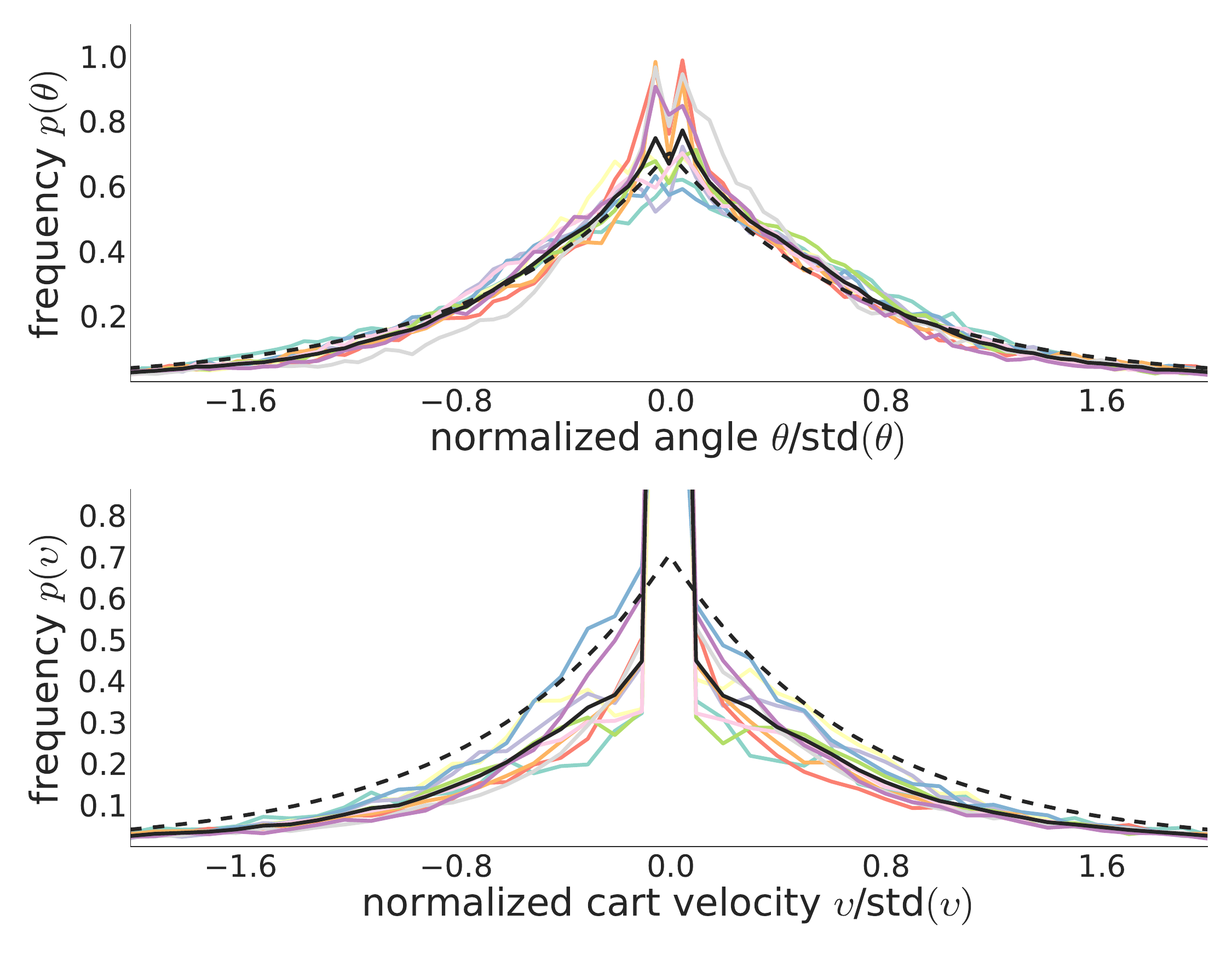} 
\caption{Fast stick: linear scale, zoom at the central part of the distribution}
\label{fig:exp_distr_fast}
\end{subfigure}
\quad
\begin{subfigure}{0.47\textwidth}
\includegraphics[width=\textwidth]{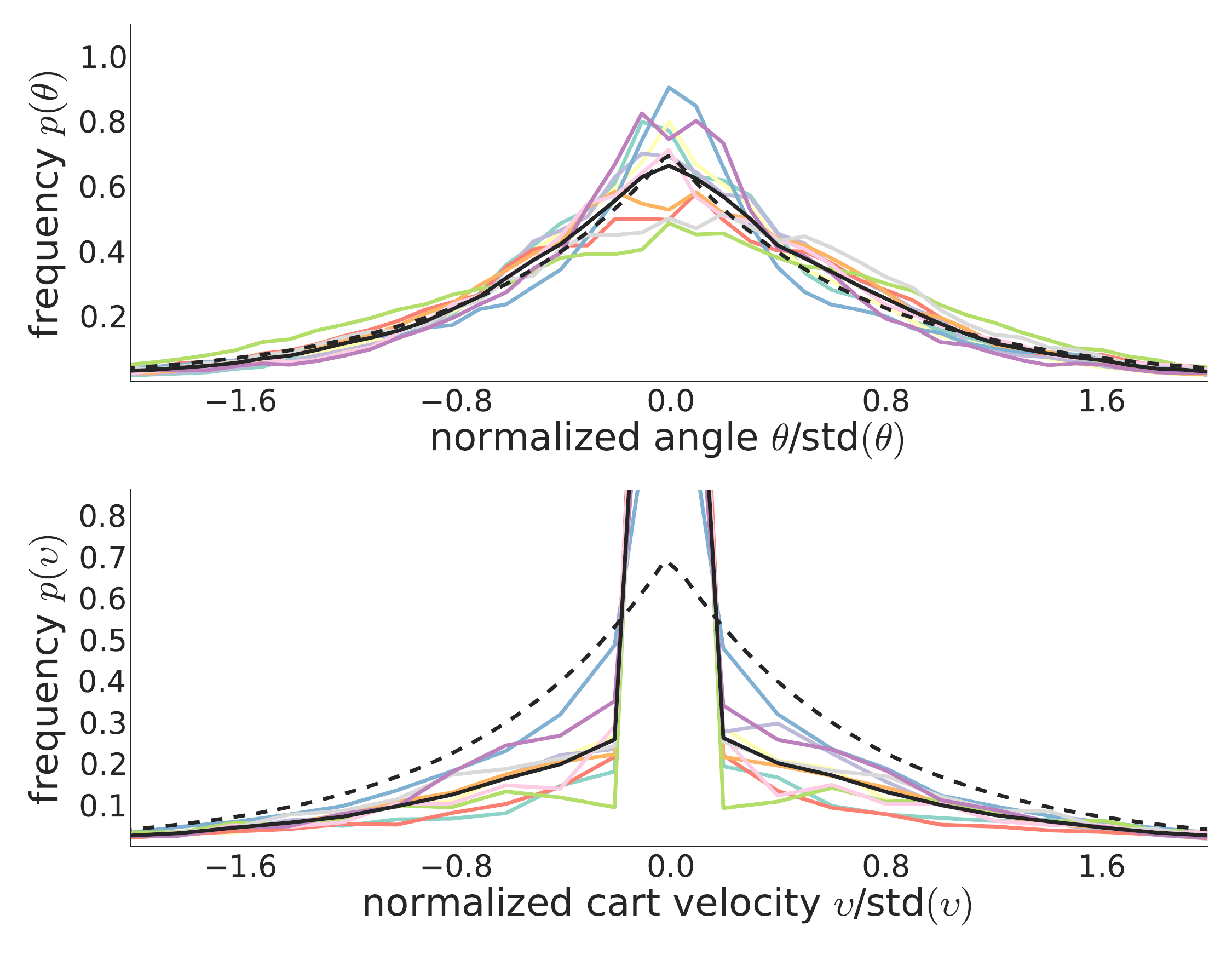} 
\caption{Slow stick: linear scale, zoom at the central part of the distribution}
\label{fig:exp_distr_slow}
\end{subfigure}
\caption{
Experimentally obtained distributions of stick angle and cart velocity. Colored lines represent the distributions for each subject. Solid black lines represent the average distributions calculated based on the aggregated data for all the subjects. Dashed lines represent the Laplace distributions (zero mean, unit variance) for reference.
}
\label{fig:exp_distr}
\end{figure*}

Based on the phase trajectories it is easy to reconstruct the typical pattern of the observed operator behavior. Given that the initial deviation of the stick from the vertical position is small, the operator halts the control so the stick falls on its own. Then, the operator takes the control over the system, moving the cart to compensate for the deviation. The corrective movements are generally imprecise, however, occasionally the operator returns the stick to a close vicinity of the upright position. Substantial errors are often corrected straight away, without waiting for the current movement to finish, which results in the multimodal fragments of the velocity profile (Fig.~\ref{fig:exp_velocity}). On the contrary, in case of moderate error the operator usually halts the control for some time after the initiated cart movement is completed, even if the resulting deviation from the upright position is evident.

Assuming the operator's response is event-driven, we analyzed the angle values corresponding to the moments when the operator starts actively controlling the system. Appealing to the studies on car following~\cite{todosiev1963action,wagner2003empirical}, we call such values the \textit{action points}. The distribution of action points is unimodal for five least skilled subjects, and bimodal for five most skilled balancers (Fig.~\ref{fig:exp_ap_distr}). This prompts that the unskilled participants attempted to react to all the detected deviations regardless of their magnitude, whereas the more competent subjects often neglected perceptible, yet still small stick deviations. This in turn prompts that the action points are determined not by the operator's limited perception abilities, but rather by the particular control strategy adopted by the operator. Notably, the distribution of action points decays exponentially regardless the subject's skill (Fig.~\ref{fig:exp_ap_distr}, right frame), indicating a 
relatively high probability of the action points corresponding to large deviations. This provides evidence against the noise-affected threshold-driven activation mechanism, which would presumably lead to the normal action point distribution centered at the hypothetical threshold value. 

To check whether the diversity of the subjects in terms of performance leads to the fundamentally different properties of the task dynamics, we analyzed the statistical distributions of the stick angle $\theta$ and the cart velocity $\upsilon$. In both conditions, both distributions are similar for all ten subjects regardless of their balancing skill (Fig.~\ref{fig:exp_distr}). In the fast stick condition the stick angle has approximately Laplacian distribution. However, the angle distribution is bimodal with a narrow gap (width of order 0.1 $\textrm{std}(\theta)$) for all the participants except Subjects 1 and 5 (Fig.~\ref{fig:exp_distr_fast}). The cart velocity distribution has a sharp peak at the origin, which corresponds to high values of $\%_{\textrm{drift}}$ and may serve as a shortcut for detecting intermittency of human control. 

In the slow stick condition the angle distribution is unimodal for all the participants and its tails are less heavy than in the fast stick condition. Otherwise, both the angle and cart velocity distributions are alike (up to scale) in the slow and fast stick conditions. The remarkable similarity of the distributions may indicate that all the subjects employ the same nonlinear mechanisms in controlling the stick in both effortful (fast condition) and relaxed (slow condition) regimes. 

\section{Model}
For simplicity, prior to elaborating the model of human control in balancing the overdamped stick, we linearize Eq.~\eqref{eq:stick} near the vertical position $\theta=0$,
\begin{equation}
 \tau \dot \theta = \theta - \frac{\tau}{l} \upsilon.
\label{eq:stick_lin}
\end{equation}

\subsection{Model construction}
Hypothetically, the stick dynamics can be described by the first-order dynamical system~\eqref{eq:stick_lin} if only the cart velocity $\upsilon$ is specified as a function of time $t$ or stick angle $\theta$. However, $\upsilon$ is actually controlled by the human operator, so it possesses its own, complex dynamics. To be able to capture this dynamics, we extend the physical phase space of the overdamped stick by a separate phase variable characterizing the actions of the operator~\cite{zgonnikov2014extended}. We thus have to specify the governing equation for the cart velocity $\upsilon$.

The experimental results reveal two distinct phases of human control, passive and active. Similarly to Bottaro et al.~\cite{bottaro2008bounded}, we hypothesize that different control mechanisms are employed in each of these phases. On one hand, during the passive control phase the operator monitors the deviation of the stick from the goal and eventually decides when to switch to the active phase. The transition from the passive to the active phase is governed by the \textit{``when-to-react'' mechanism} (control activation). On the other hand, during the active control phase the stick is returned to some vicinity of the vertical position by the corrective action of the operator, which is implemented by the \textit{``how-to-react'' mechanism} (control execution). Within this two-mechanism framework we hypothesize that the ``how-to-react'' mechanism generates corrective movements of open-loop type, and the ``when-to-react'' mechanism implements noise-driven control activation. 

Human control is often characterized by open-loop, preprogrammed corrective actions, rather than closed-loop feedback strategies~\cite{loram2002human, loram2005human, ben2008minimum, gawthrop2011intermittent}. In the current context it implies that once the operator launches a hand movement to compensate for the detected stick deviation, this movement is not interrupted until fully executed. Unfortunately, despite the currently gained understanding of the open-loop properties of human control, the corresponding mathematical formalism still has to be developed. For this reason the present model mimics the experimentally observed dynamics by utilizing a zeroth-order, continuous feedback approximation to the presumably open-loop trajectories of the system in the active phase.

The continuous approximation to open-loop control is built around the assumption that the operator behavior in driving the stick towards the vertical angle is optimal in some sense. Particularly, when compensating for a stick deviation, the operator supposedly chooses the response $\upsilon$ in a way to minimize the loss function based on the measure
\begin{equation}
F(\upsilon,\dot \upsilon,\theta) = \frac{\tau^2}{2l^2}\left(\upsilon^2 + \tau_m^2\dot{\upsilon}^2\right) + \frac{\theta^2}{2\theta_m^2}
\label{eq:loss}
\end{equation}
where $\theta_m$ and $\tau_m$ are non-negative constant parameters. The actions of the operator are then described by the linear feedback (see Appendix~\ref{app:openloop} for details)
\begin{equation}
\dot \upsilon = \alpha l \theta - \beta \upsilon,\quad \alpha = \beta^2/2,\,\beta>0.
\label{eq:control_lin}
\end{equation}
We use Eq.~\eqref{eq:control_lin} to mimic the dynamics of the operator-controlled cart during the active phase.

The pivot point of the present model is that control activation is not threshold-driven (as assumed by virtually all available studies on human control), but noise-driven. We suggest that the operator decision when to react is determined by the noise-mediated interplay between two stimuli. 

On one hand, the operator is averse to actively controlling the stick; the zero value of the cart velocity, $\upsilon = 0$,  is thus attractive to the operator. Indeed, a number of possible factors (e.g., considerations of energy efficiency, or inability to precisely control the cart in compensating for small stick deviations) may cause the operator to be biased towards not moving the cart even in presence of detectable deviation. 

On the other hand, the ultimate goal, to maintain the stick upwards, inclines the operator to engage in active control over the stick. Moreover, in the absence of operator's response the angular deviation of the stick grows exponentially, presumably increasing the strength of the stimulus to act. 

The two stimuli, one inclining the operator to act, and the other one resulting in resistance to change the status quo $\upsilon = 0$, are assumed to compete stochastically. The dynamics of their interplay can be captured by modifying Eq.~\eqref{eq:control_lin} in the following way
\begin{equation}
 \dot \upsilon = \Omega(\upsilon) [\alpha l \theta - \beta \upsilon] + f(t),
\label{eq:control_general}
\end{equation}
where $f(t)$ is the random force of small amplitude and the cofactor $\Omega$ is a function of $\upsilon$ such that $\Omega(\upsilon) \approx 0$ if $\upsilon \approx 0$ and $\Omega(\upsilon) \approx 1$ otherwise. Generally, any function matching these conditions can be used; for reason of simplicity we choose the ansatz
 \begin{equation}
 \Omega(\upsilon) = \frac{\upsilon^2}{\upsilon^2 + \eta^2}, 
 \label{eq:omega}
 \end{equation}
where $\eta > 0$ is a constant parameter. 

In Eq.~\eqref{eq:control_general} the cofactor $\Omega$ reflects the attractive properties of the status quo manifold $\upsilon = 0$, whereas the cofactor $[\alpha l \theta - \beta \upsilon]$ represents the stimulus to act. The stochastic term $f(t)$ is introduced to allow for the possibility of the system's escape from the unstable manifold $\upsilon = 0$ so that the active control term, $[\alpha l \theta - \beta \upsilon]$, can eventually come into play. It is assumed to have the form
\begin{equation}
f(t) = \epsilon \xi, 
 \label{eq:noise}
\end{equation}
where $\xi$ is white Gaussian noise and $\epsilon \ll 1$ is the noise amplitude. We wish to underline that the random force $f(t)$ does not represent the sensorimotor noise, but instead serves to mimic the stochasticity of the operator's decision when to react. 

Both components of the proposed two-mechanism framework reflect complex cognitive operations which take time in the real control process. However, in case of overdamped stick balancing introducing delay in the model~\eqref{eq:stick_lin},\eqref{eq:control_general},\eqref{eq:omega},\eqref{eq:noise} would not change its basic dynamics. Indeed, during the time required for the two mechanisms to process the detected deviation $\theta (t_0)$ this deviation increases by a factor depending on the response delay $\Delta$ and the time scale of the uncontrolled stick motion $\tau$. Given $\upsilon=0$, the solution of the initial value problem for Eq.~\eqref{eq:stick_lin} yields $$\theta (t_0+\Delta)=\theta (t_0) e^{\Delta/\tau}.$$ Consequently, as long as $\Delta/\tau$ remains small enough, the delay in the operator's response has minor impact on the stick dynamics, affecting only the amplitude of the stick oscillations.

\subsection{Model dynamics}
\begin{figure*}
\centering
\begin{subfigure}{0.4\textwidth}
\includegraphics[width=\textwidth]{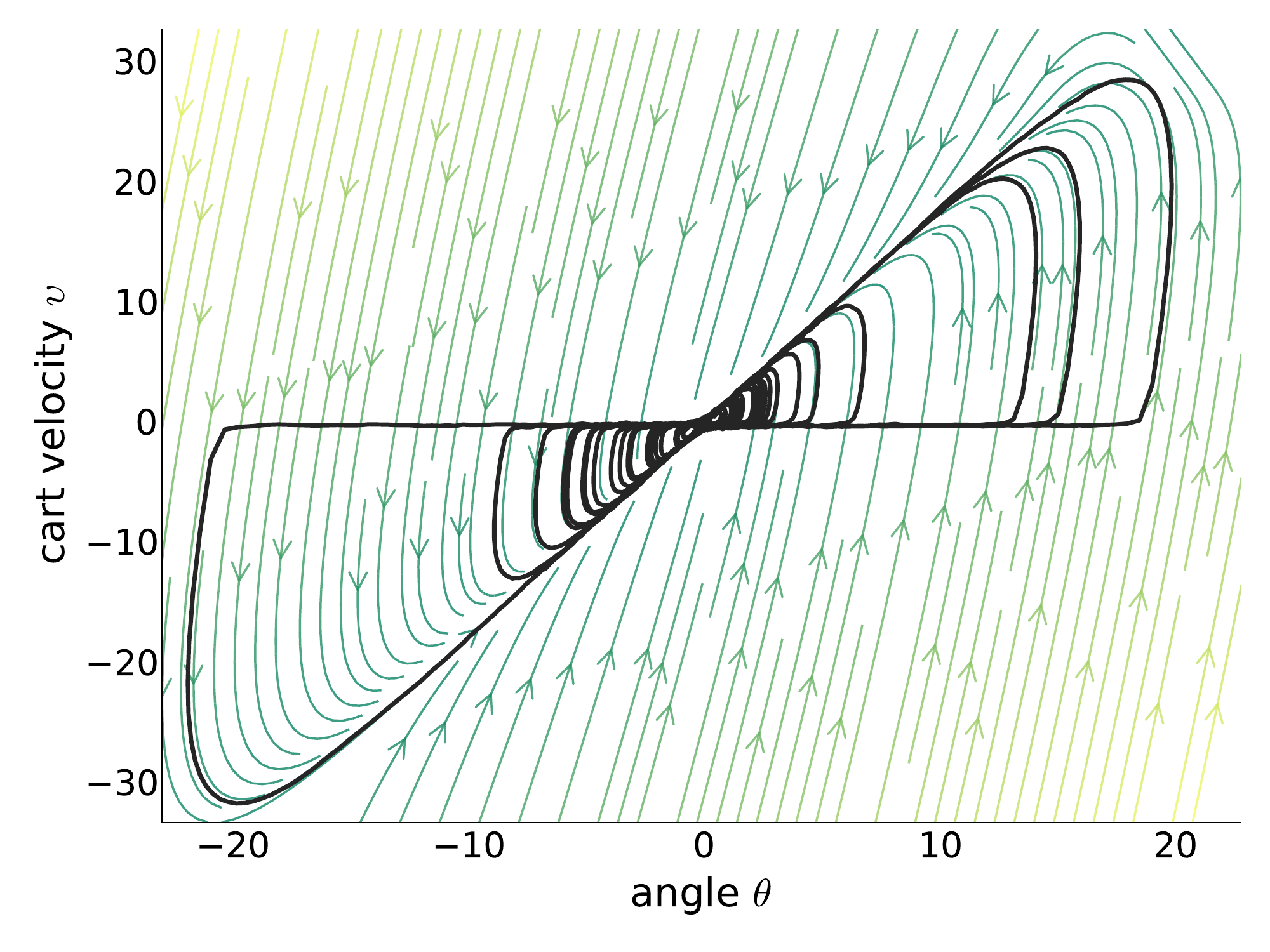} 
\caption{$\gamma = 12,\,\sigma=7.5,\,\varepsilon = 0.2$}
\label{fig:sim_phase_node}
\end{subfigure}
\quad
\begin{subfigure}{0.4\textwidth}
\includegraphics[width=\textwidth]{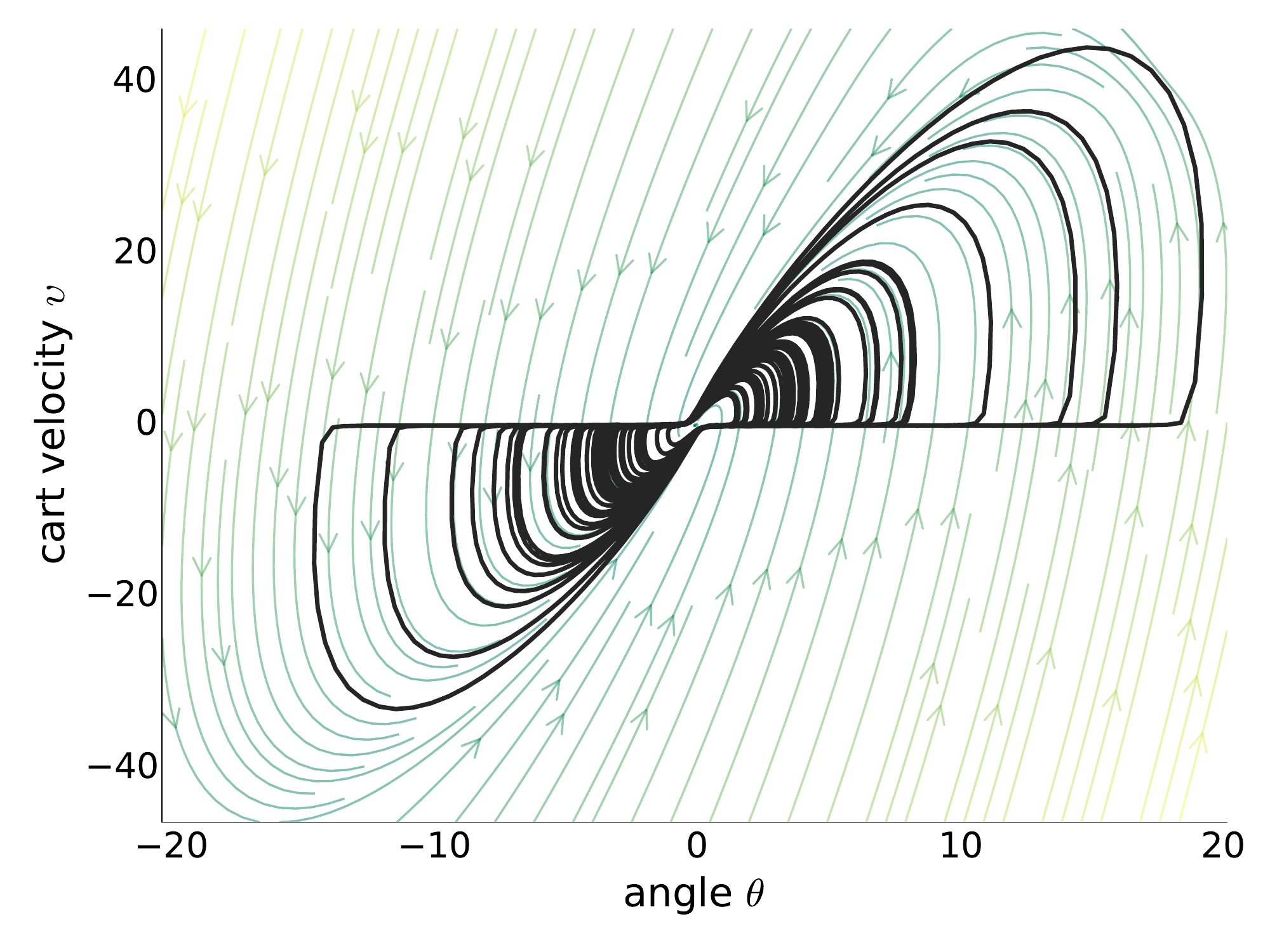} 
\caption{$\gamma = 22,\,\sigma=7.5,\,\varepsilon = 0.2$}
\label{fig:sim_phase_focus_gamma_low}
\end{subfigure}
\\ 
\begin{subfigure}{0.4\textwidth}
\includegraphics[width=\textwidth]{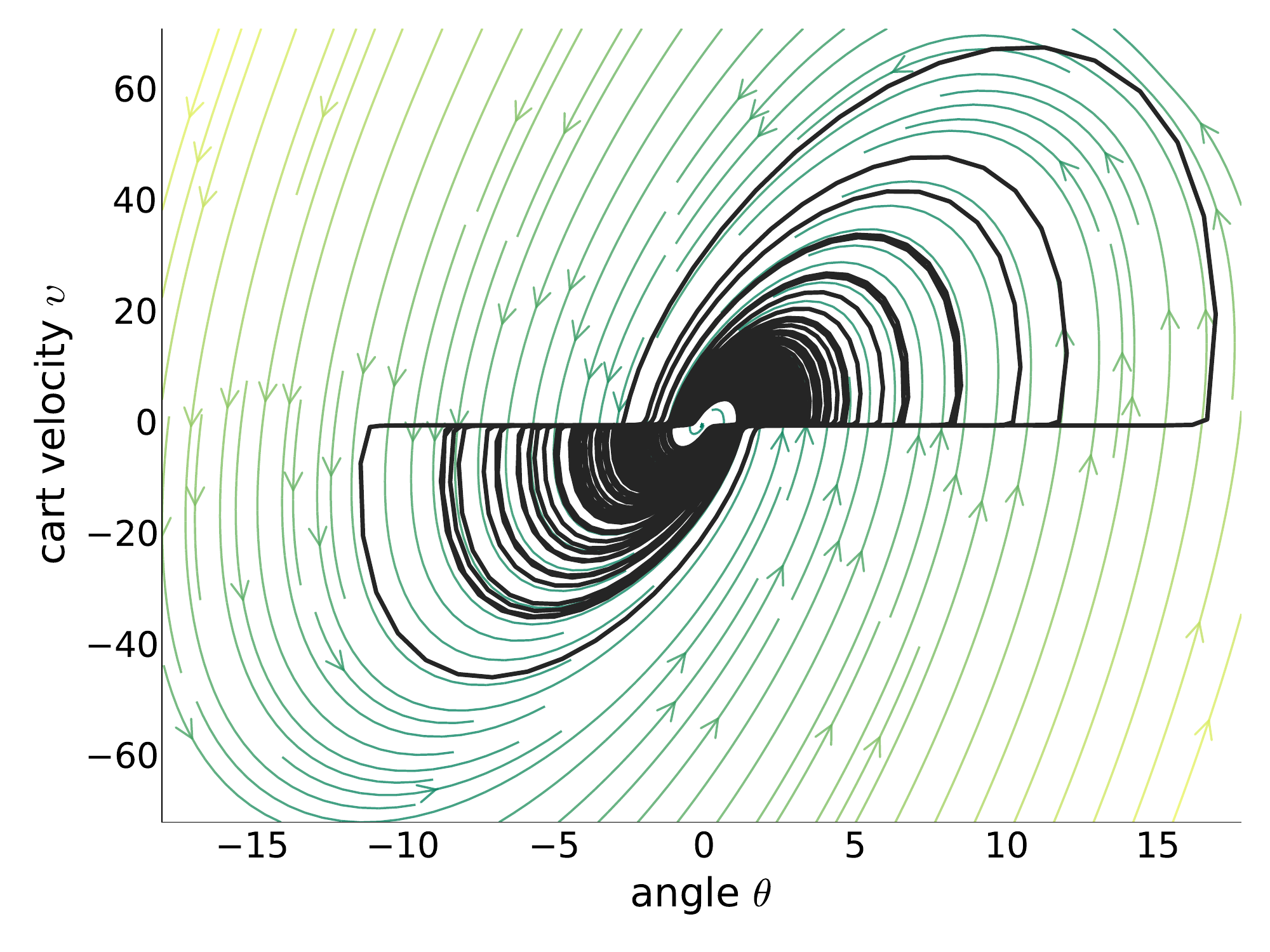} 
\caption{$\gamma = 48,\,\sigma=7.5,\,\varepsilon = 0.2$}
\label{fig:sim_phase_focus_gamma_high}
\end{subfigure}
\quad
\begin{subfigure}{0.4\textwidth}
\includegraphics[width=\textwidth]{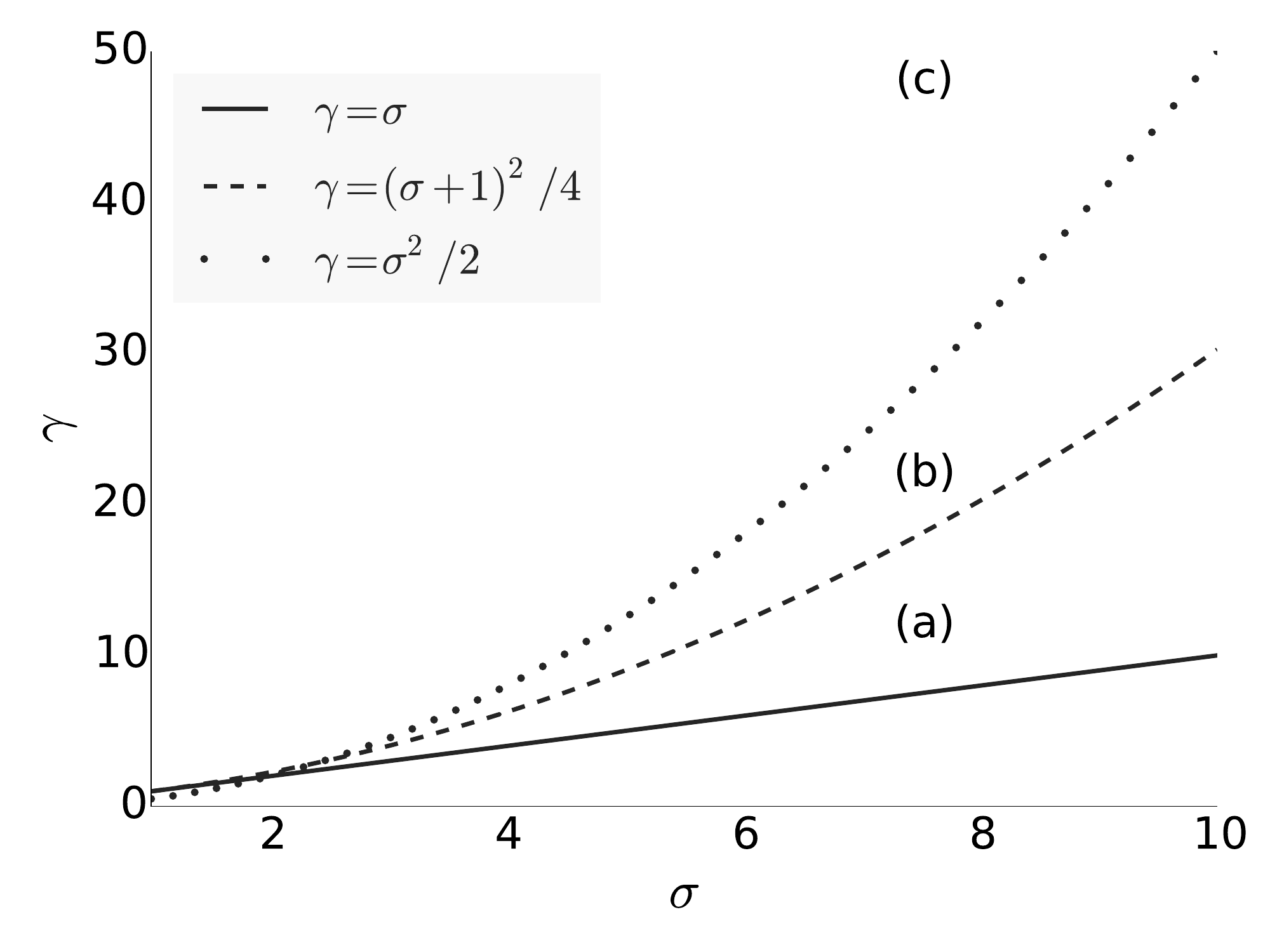} 
\caption{Phase diagram}
\label{fig:sim_phase_diagram}
\end{subfigure}
\caption{
Typical dynamics of the model~\eqref{eq:model}. Frames (a)-(c) show the phase trajectories (solid black lines) generated by the system~\eqref{eq:model}. Faint arrowed lines represent the force field of the linear system~\eqref{eq:model_linear}. Frame (d) presents the phase diagram of the model~\eqref{eq:model_linear} in the $\sigma\gamma$ plane. The solid line represents the stability boundary, the dashed line marks the border between the node-type and focus-type dynamics and the dotted line corresponds to the feedback optimality condition. Dynamics of the system in each region of the diagram are illustrated in the corresponding frame.
}
\label{fig:sim_phase}
\end{figure*}

Prior to analyzing the dynamics of the model, we rescale the variables 
\begin{equation*}
\label{eq:rescale}
 t \to t \tau,\,\, \theta \to \theta \eta \tau / l,\,\, \upsilon \to \upsilon \eta, 
\end{equation*}
so that in new, dimensionless variables $t,\,\theta,\,\upsilon$ the model~\eqref{eq:stick_lin},\eqref{eq:control_general},\eqref{eq:omega},\eqref{eq:noise} takes the form
\begin{equation}
\label{eq:model}
 \begin{aligned}
  \dot \theta & = \theta - \upsilon, \\
  \dot \upsilon & = \Omega(\upsilon) [\gamma \theta - \sigma \upsilon] + \varepsilon \xi,\\
 \Omega & (\upsilon) = \frac{\upsilon^2}{\upsilon^2 + 1},
 \end{aligned}
\end{equation}
where $\gamma = \alpha \tau^2$, $\sigma = \beta \tau$, and $\varepsilon = \epsilon \sqrt{\tau}/\eta$. Parameter $\eta$ thus has no impact on the core dynamics of the original model, just defining the scale of the system motion. The necessary condition for the feedback~\eqref{eq:control_lin} to be optimal, $\alpha = \beta^2/2$, takes the form $\gamma = \sigma^2/2$. For reasons of flexibility, however, we consider the parameters $\gamma$ and $\sigma$ to be independent in the general case.

Typical phase trajectories exhibited by the model~\eqref{eq:model} are represented in Figs.~\ref{fig:sim_phase_node}, \ref{fig:sim_phase_focus_gamma_low}, \ref{fig:sim_phase_focus_gamma_high}. The initially perturbed system moves along the $\theta$-axis with the cart velocity $\upsilon$ close to zero, so that $\dot \theta \approx \theta$. This motion regime represents the passive control phase. As the angle $\theta$ increases, the system may escape from the vicinity of the manifold $\upsilon=0$ due to the random force $\varepsilon \xi$. Small fluctuations of the system moving along the axis $\upsilon = 0$ result, sooner or later, in the situation when the trapping effect of $\Omega$ is suppressed by the growing magnitude of the cofactor $[\gamma \theta - \sigma \upsilon]$. This triggers the sharp transition from $\dot \upsilon \approx 0$ to $\dot \upsilon \approx \gamma \theta - \sigma \upsilon$, i.e., the transition from the passive to the active control phase. However, in case the random force is absent, $\varepsilon = 0$, the system steadily moves away from the equilibrium along the $\theta$-axis. The switching from the passive to the active phase is thus driven solely by noise. In what follows we first explore the system properties assuming $\varepsilon = 0.2$. After that, we examine how the noise intensity affects the system behavior.

The dynamics of the system~\eqref{eq:model} in the active control phase are defined by the linear system 
\begin{equation}
\label{eq:model_linear}
 \begin{aligned}
  \dot \theta & = \theta - \upsilon, \\
  \dot \upsilon & = \gamma \theta - \sigma \upsilon,
 \end{aligned}
\end{equation}
except the vicinity of the $\theta$-axis, where the effect of the cofactor $\Omega$ becomes essential. Namely, when $\upsilon$ approaches zero, the trajectory of the system~\eqref{eq:model} smoothly adjoins the $\theta$-axis, i.e., the system switches back to the passive phase instead of being driven precisely to the equilibrium.

In the passive control phase, $\upsilon \approx 0$, the system~\eqref{eq:model} is unstable, $\dot \theta \approx \theta$. Thus, in order for the system motion to be overall bounded, the absolute value of the stick angle should decrease as an outcome of the single active correction: $(\theta=\theta_{\text{start}},\,\upsilon=0) \to (\theta=\theta_{\text{end}},\,\upsilon=0)$, $|\theta_{\text{end}}| < |\theta_{\text{start}}|$. During the active phase, first, the effect of the random force is minor, and, second, the system dynamics is essentially linear. Therefore, the stability of the system~\eqref{eq:model_linear} is the necessary condition for the dynamics of the system~\eqref{eq:model} to be bounded. This requires
 \begin{equation}
  \label{eq:plausible_params}
   \sigma > 1,\quad \gamma > \sigma.
 \end{equation}
Within the assumption~\eqref{eq:plausible_params}, the particular values of parameters $\gamma$ and $\sigma$ define, first, the form of the system trajectory, and, second, the time scale of the system motion in the active phase. 

As long as $\gamma < (\sigma+1)^2/4$, the linear system~\eqref{eq:model_linear} has stable equilibrium of the node type at the origin. In this case the trajectory of the system~\eqref{eq:model} practically reaches the origin as a result of each active phase (Fig.~\ref{fig:sim_phase_node}). In contrast, in case of focus-type active phase dynamics, $\gamma > (\sigma+1)^2/4$, the system switches to the passive phase at the non-zero angles (Figs.~\ref{fig:sim_phase_focus_gamma_low},~\ref{fig:sim_phase_focus_gamma_high}), which more resembles the experimentally observed behavior. In what follows we consider only the latter case, discarding the case of node-type dynamics as less physically plausible. 

Importantly, for matter of convenience we also stick to the case of optimal feedback, 
 \begin{equation}
  \label{eq:opt_params}
   \gamma = \sigma^2/2.
 \end{equation}
Due to linearity of the system behavior in the active phase, departures from the optimality condition~\eqref{eq:opt_params} do not considerably affect the results of the further analysis, which has also been verified numerically.

In case of focus-type dynamics the duration $T$ of the active phase fragments practically does not depend on the initial deviation $\theta_0$. Indeed, solving the boundary value problem for the linear system~\eqref{eq:model_linear},\eqref{eq:opt_params} and the boundary conditions $\theta(0)=\theta_0,\,\upsilon(0)=0$, $\theta(T)=\theta_T,\,\upsilon(T)=0$ with respect to unknown time $T$, we get
\begin{equation}
\label{eq:act_ph_dur}
 T = 2\pi\Big(\big(\sigma-1\big)^2-2\Big)^{-1/2}.
\end{equation}
Experimentally obtained values of $T$ are of order unity. Hence, the results of the further analysis are verified for $\sigma$ corresponding to $T\sim1$, and are illustrated for $\sigma=3.5$, $\sigma = 7.5$, and $\sigma = 13.6$ (matching $T=3.0$, $T=1.0$, and $T=0.5$, correspondingly).

\begin{figure}[!ht]
\centering
\begin{subfigure}{0.9\columnwidth}
\includegraphics[width=1.0\textwidth]{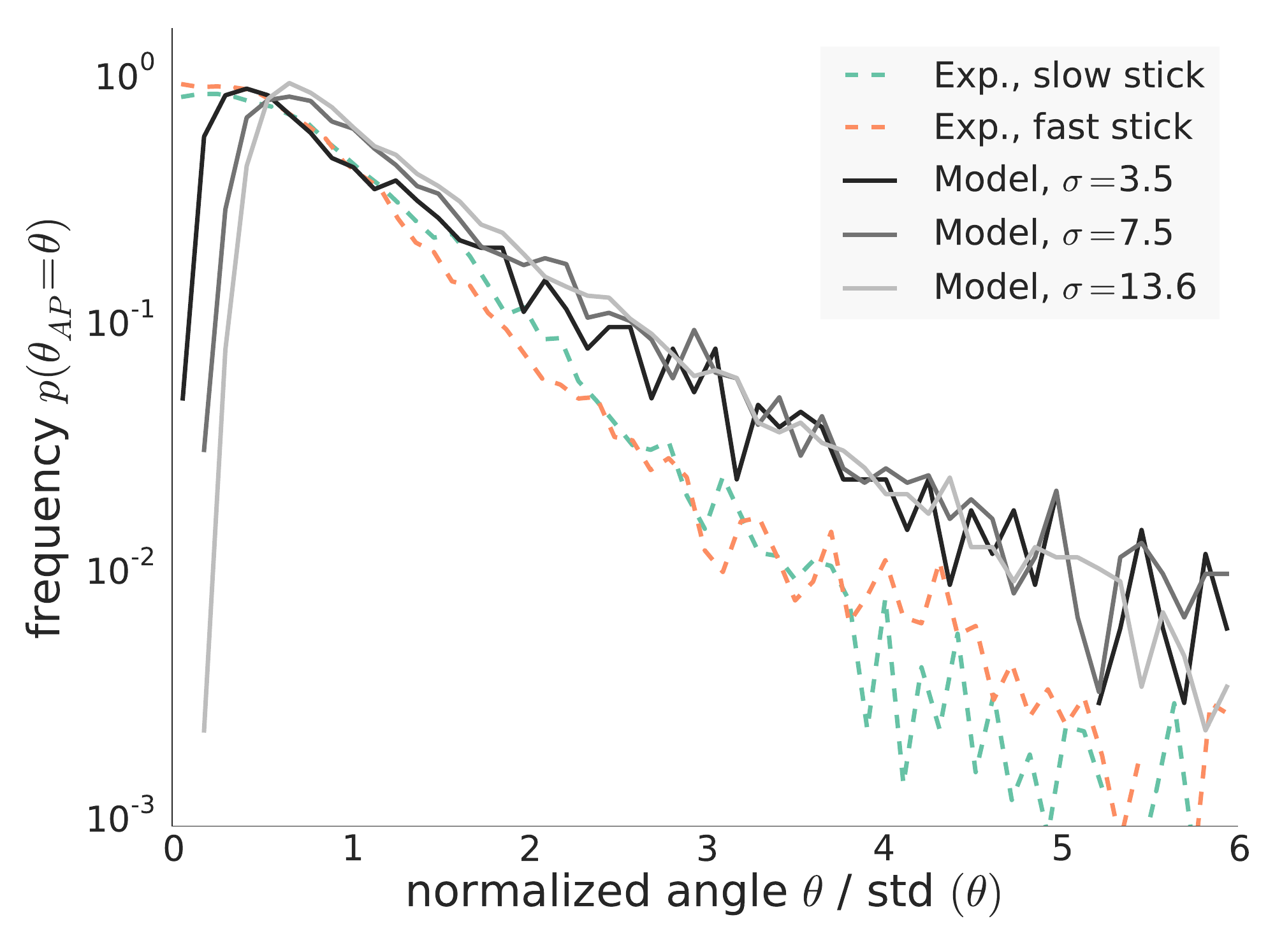}
\caption{Action point distribution}
\label{fig:sim_ap_distr}
\end{subfigure}
\\
\begin{subfigure}{1.0\columnwidth}
\includegraphics[width=1.0\columnwidth]{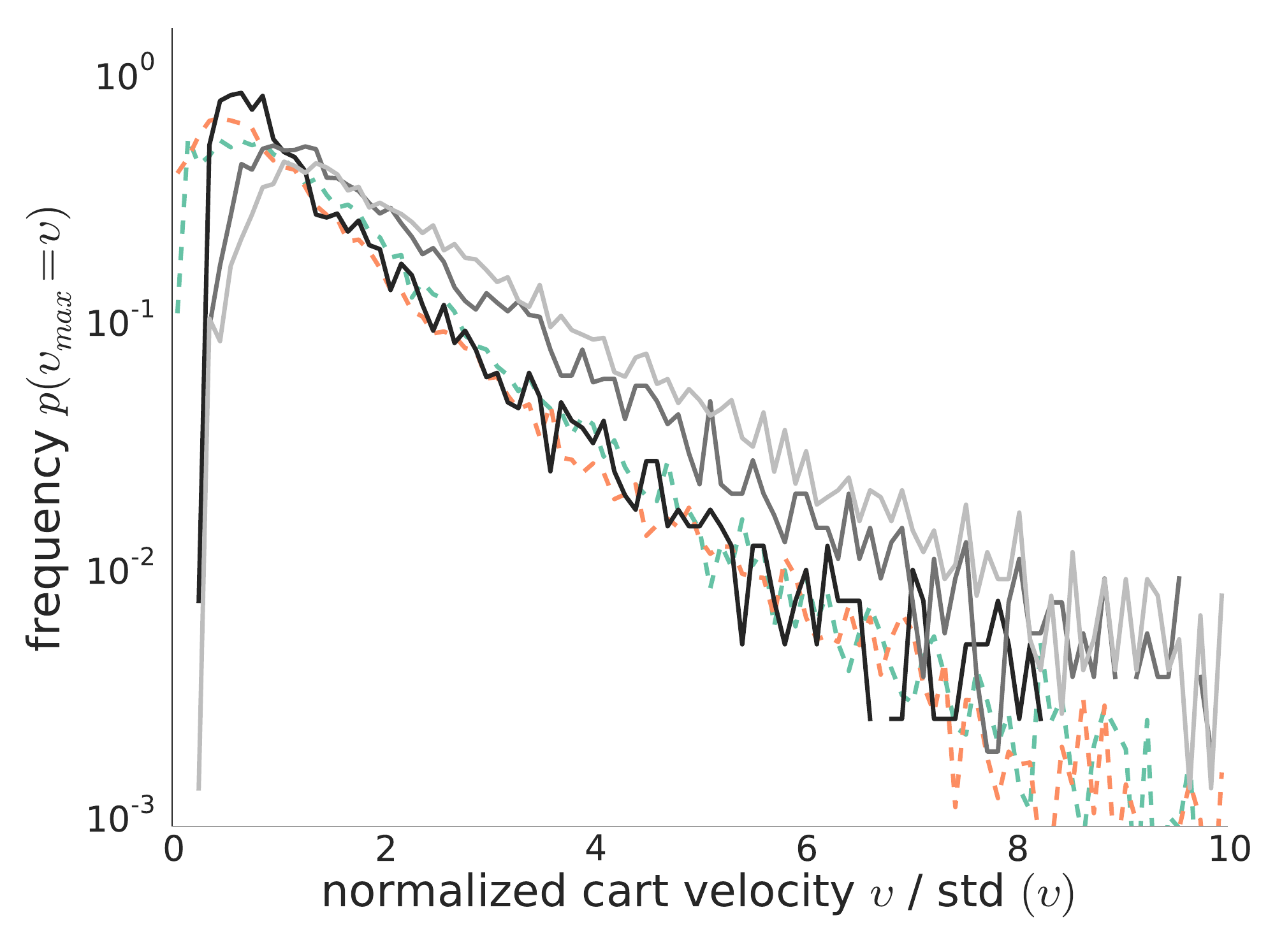}
\caption{Peak velocity distribution.}
\label{fig:sim_vmax_distr}
\end{subfigure}
\\
\begin{subfigure}{1.0\columnwidth}
\includegraphics[width=1.0\columnwidth]{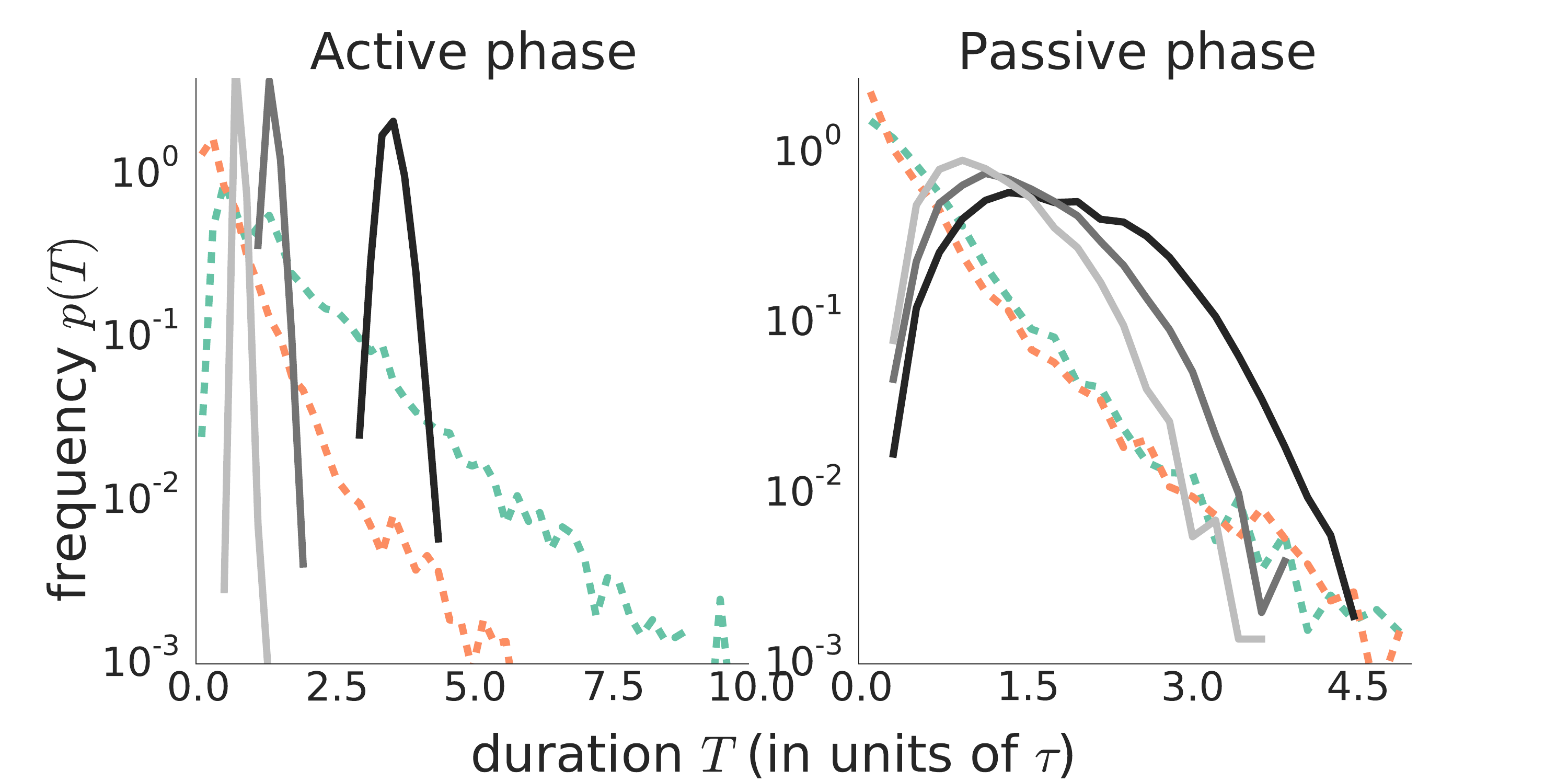}
\caption{Phase duration distribution}
\label{fig:sim_ph_dur}
\end{subfigure}
\caption{
Statistical distributions exhibited by the system~(\ref{eq:model}) for $\gamma = \sigma^2/2$, $\varepsilon = 0.2$. Solid lines represent distributions for different $\sigma$. Dashed lines represent the experimentally obtained distributions averaged across all subjects. In Fig.~\ref{fig:sim_vmax_distr} all positive local maxima and negative local minima of the cart velocity $\upsilon$ are treated as peak values.
}
\label{fig:sim_ap_ph_distr}
\end{figure}

\begin{figure}
\centering
\includegraphics[width=1.0\columnwidth]{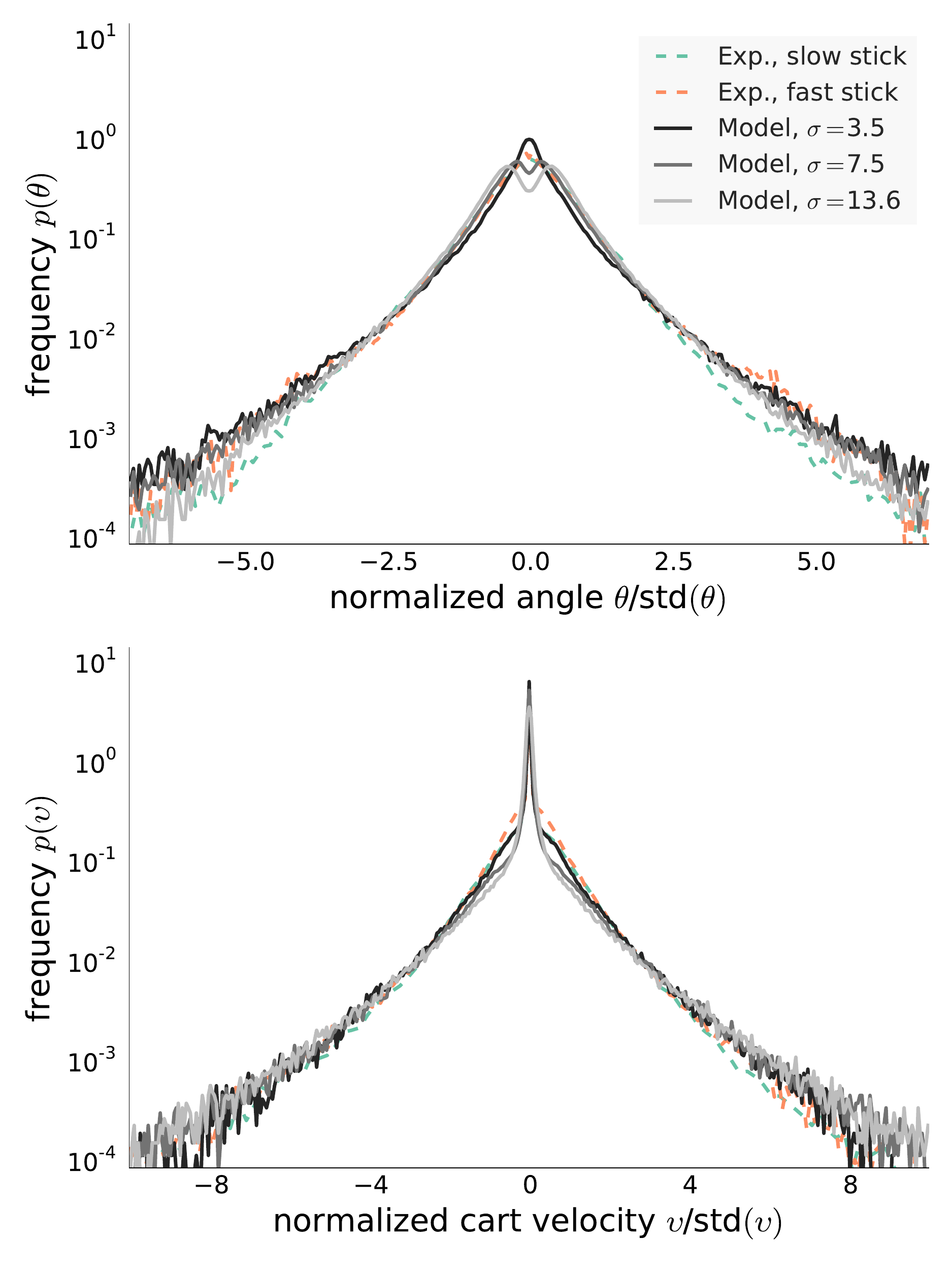}
\caption{Stick angle (top frame) and cart velocity (bottom frame) distributions exhibited by the system~(\ref{eq:model}) for different values of parameter $\sigma$. The parameter $\gamma$ was set to $\sigma ^2/2$ to match the optimality condition~\eqref{eq:opt_params}; the noise intensity was fixed, $\varepsilon = 0.2$.}
\label{fig:sim_distr_sigma}
\end{figure}

\begin{figure*}[!ht]
\centering
\begin{subfigure}{0.24\textwidth}
\includegraphics[width=\textwidth]{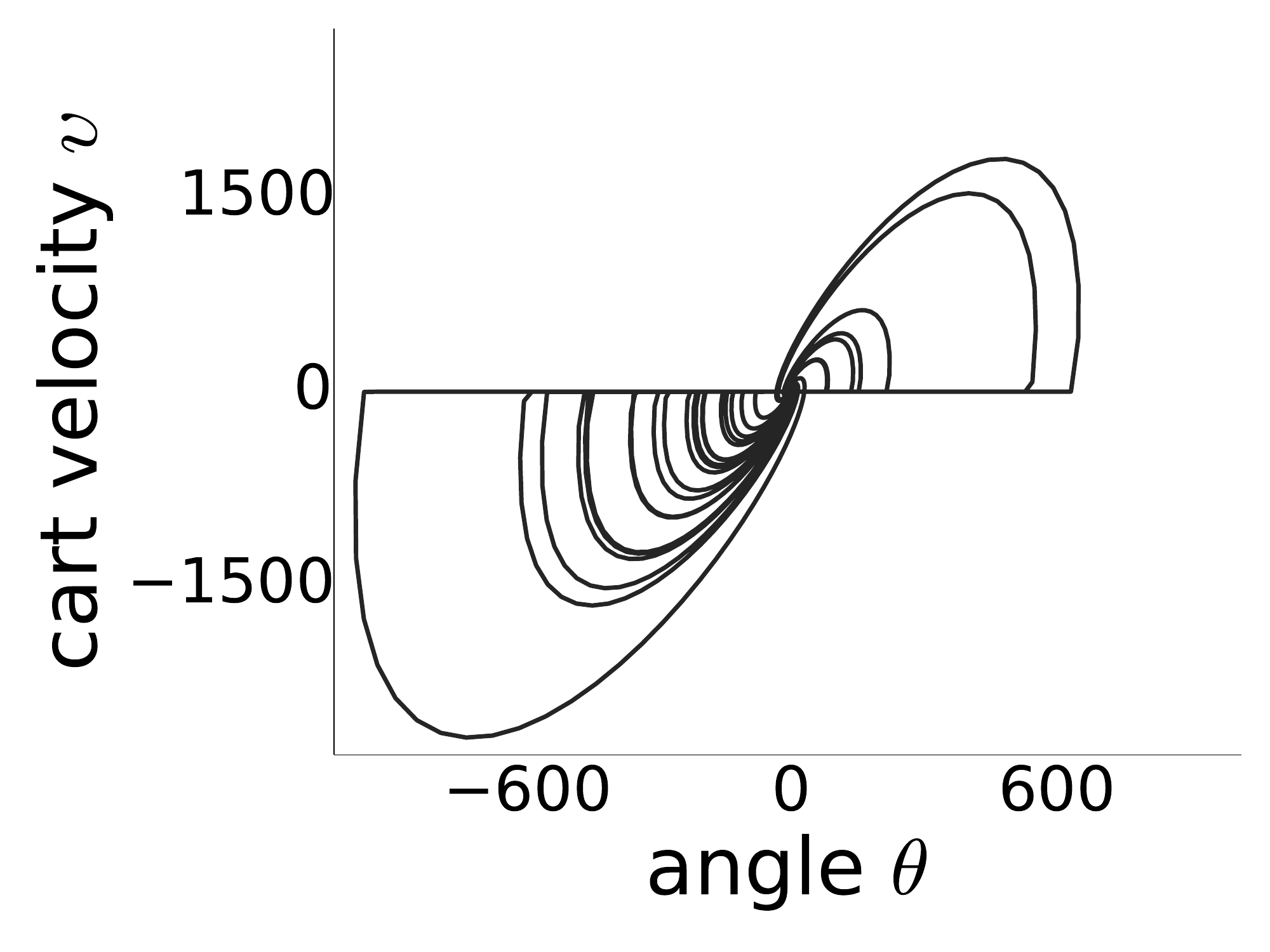} 
\caption{$\varepsilon = 0.001$}
\label{fig:sim_phase_a}
\end{subfigure}
\begin{subfigure}{0.24\textwidth}
\includegraphics[width=\textwidth]{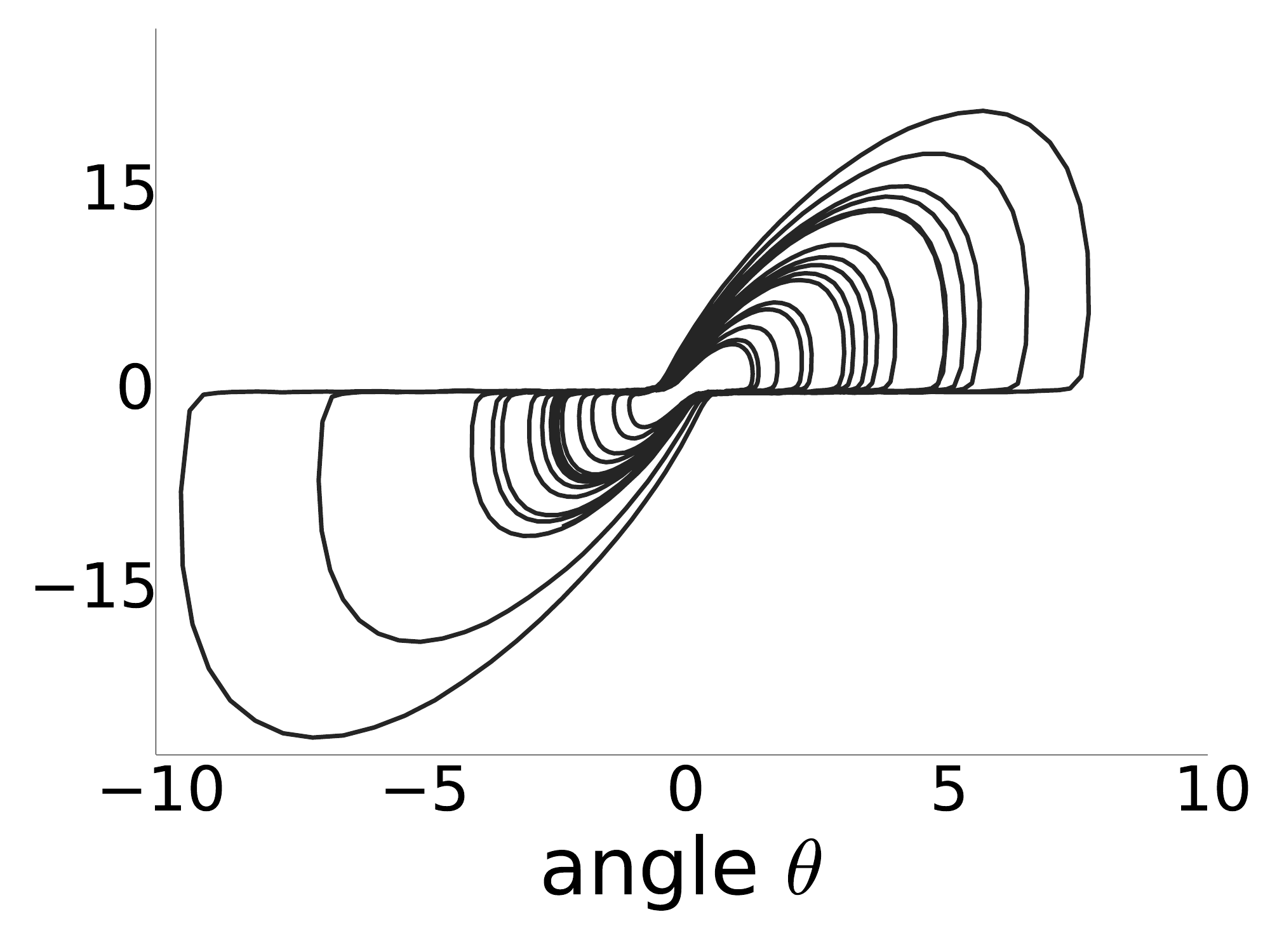} 
\caption{$\varepsilon = 0.13$}
\label{fig:sim_phase_b}
\end{subfigure}
\begin{subfigure}{0.24\textwidth}
\includegraphics[width=\textwidth]{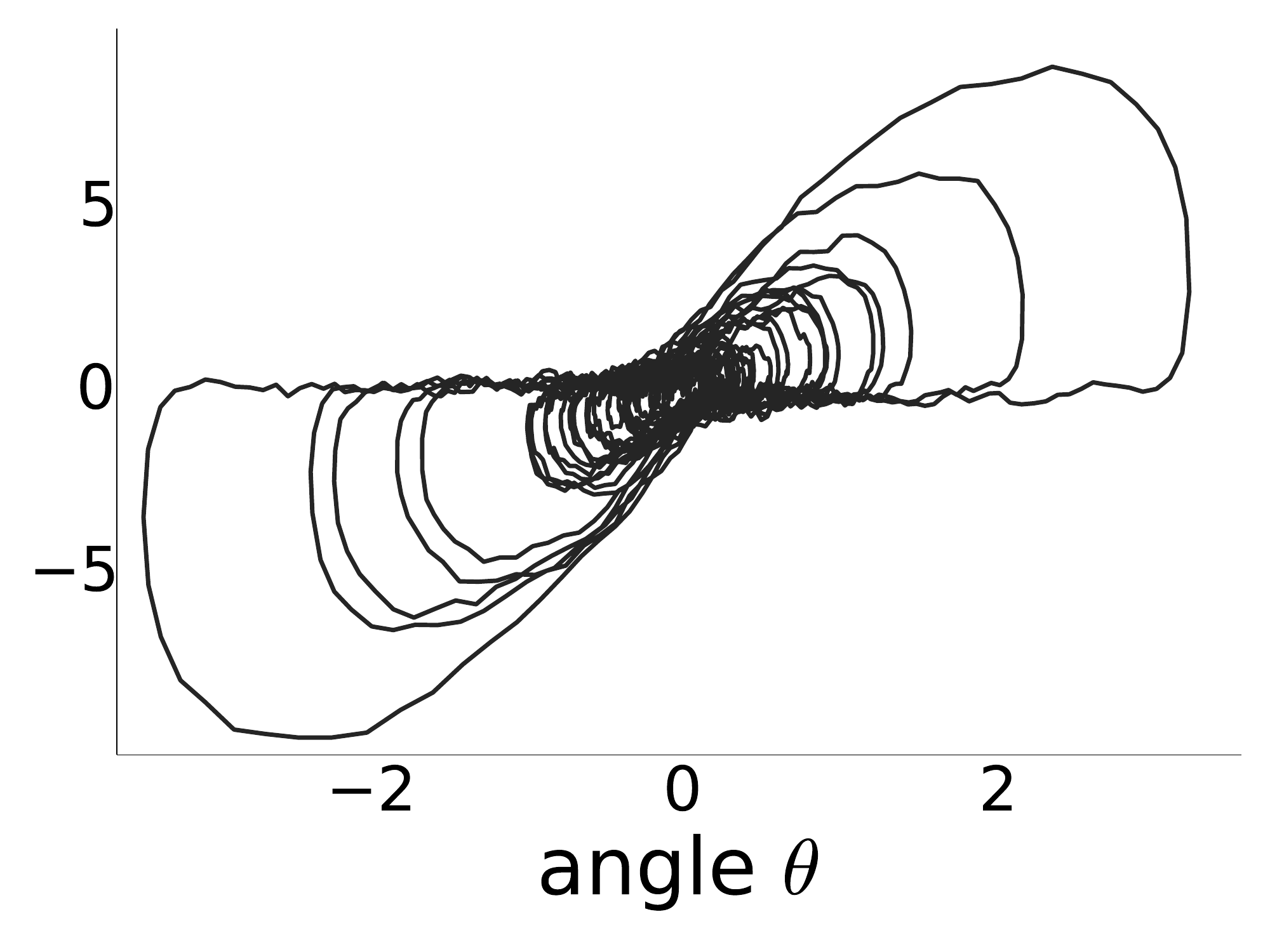} 
\caption{$\varepsilon = 0.8$}
\label{fig:sim_phase_c}
\end{subfigure}
\begin{subfigure}{0.24\textwidth}
\includegraphics[width=\textwidth]{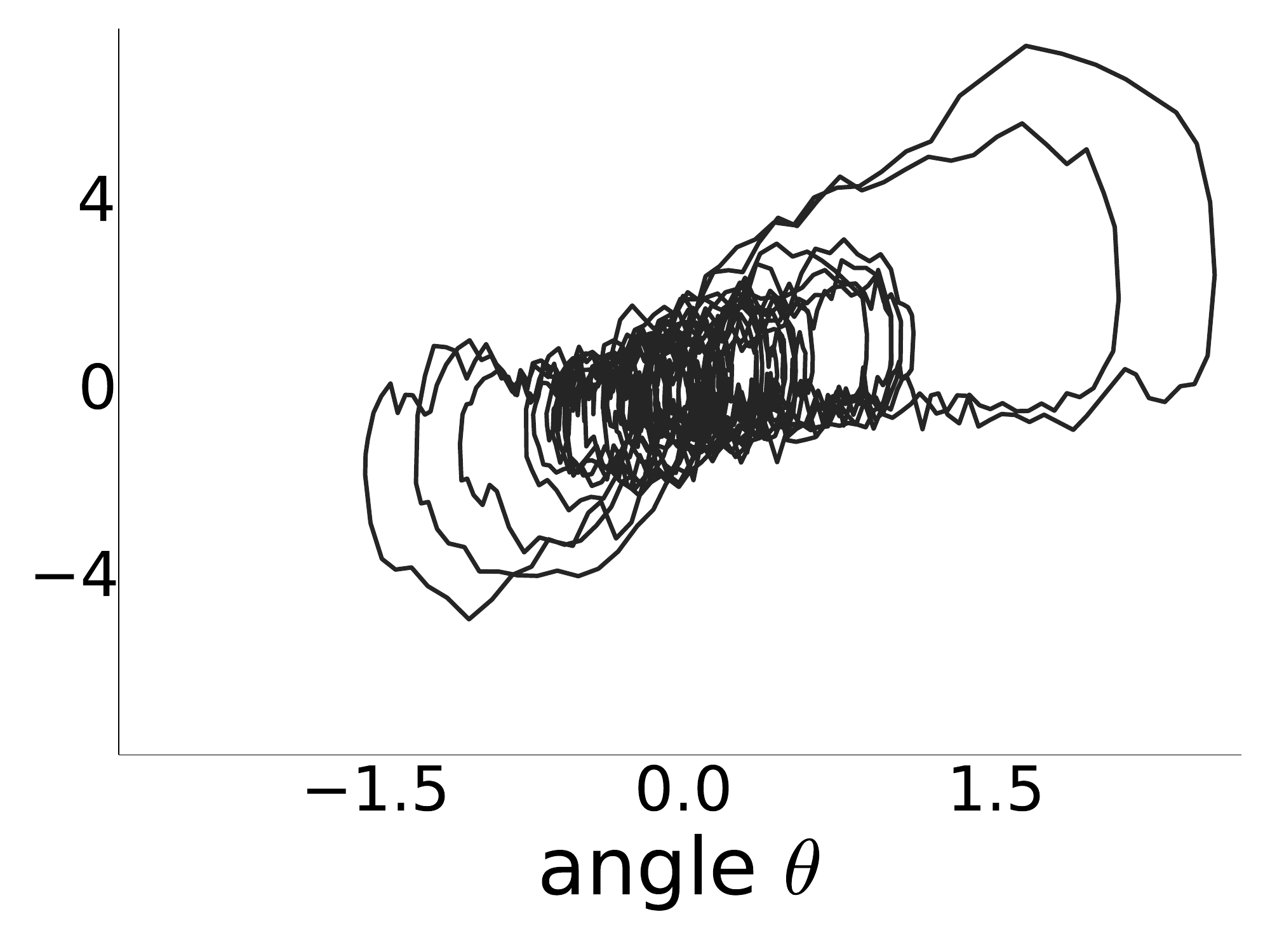} 
\caption{$\varepsilon = 2.0$}
\label{fig:sim_phase_d}
\end{subfigure}
\\ 
\begin{subfigure}{0.32\textwidth}
\includegraphics[width=\textwidth]{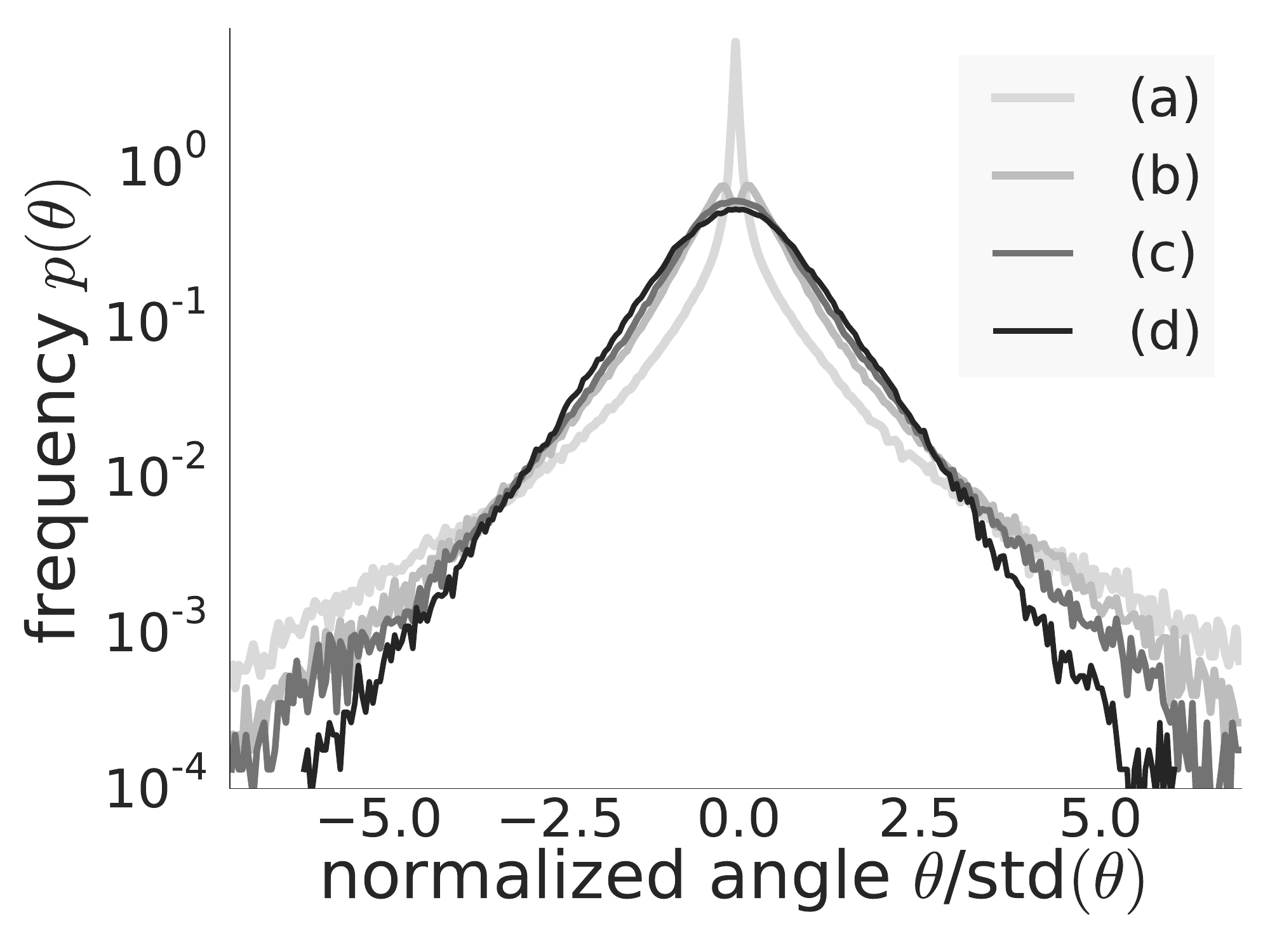} 
\caption{Angle distribution}
\label{fig:sim_angle_distr}
\end{subfigure}
\begin{subfigure}{0.32\textwidth}
\includegraphics[width=\textwidth]{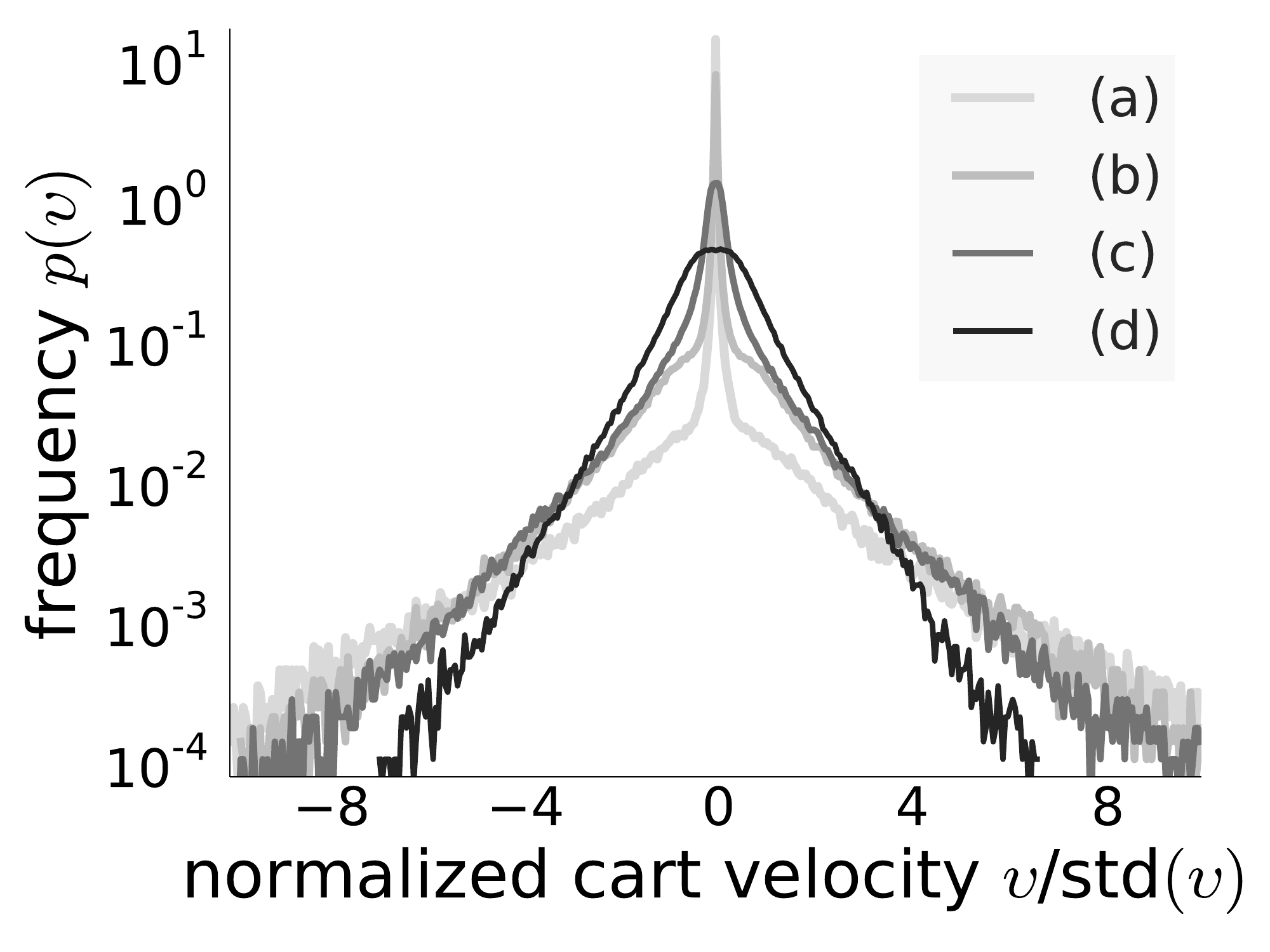} 
\caption{Velocity distribution}
\label{fig:sim_velocity_distr}
\end{subfigure}
\begin{subfigure}{0.32\textwidth}
\includegraphics[width=\textwidth]{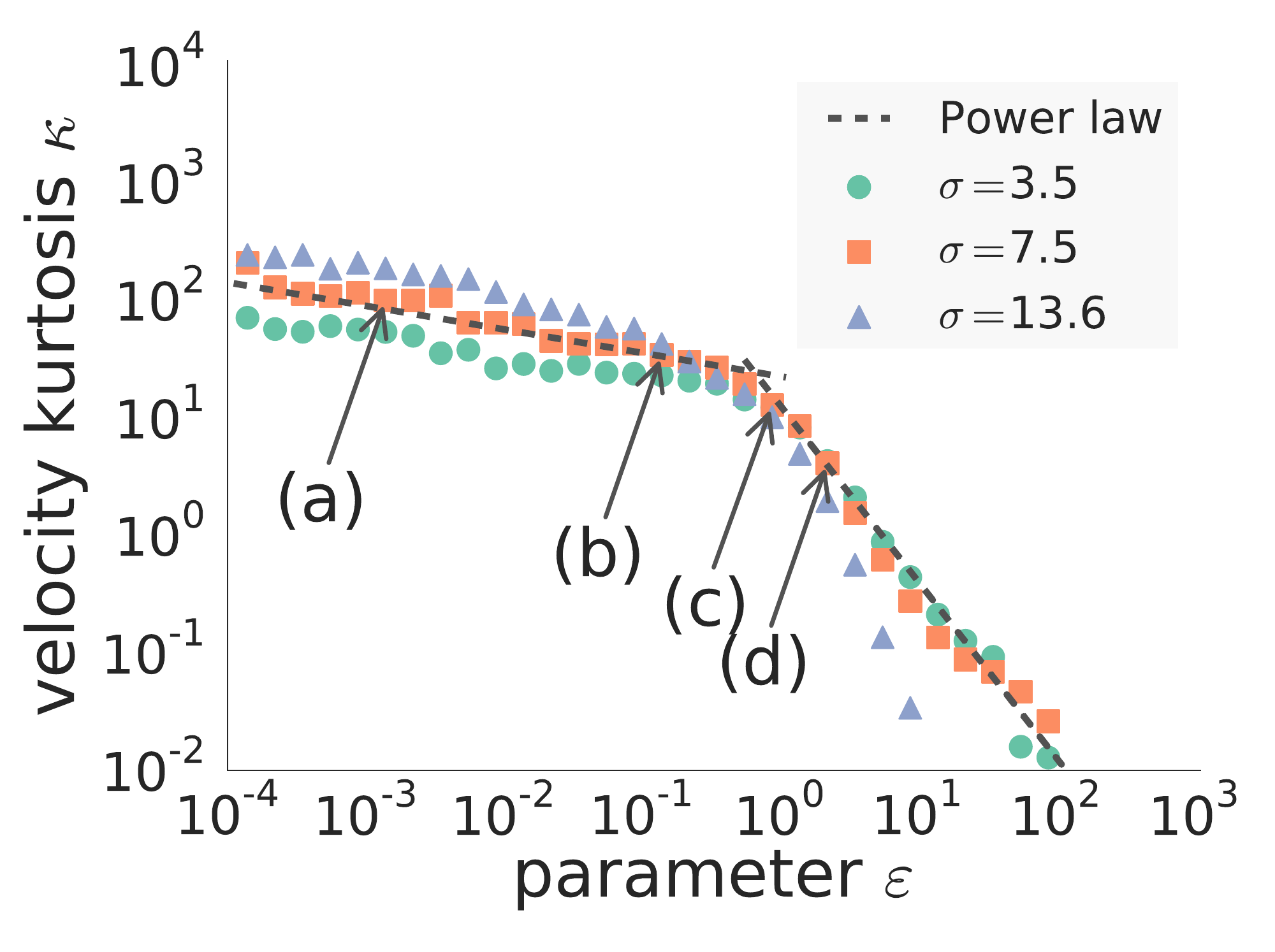} 
\caption{Velocity kurtosis}
\label{fig:sim_kurtosis}
\end{subfigure}
\caption{
Dynamics of the model~\eqref{eq:model} depending on the parameter $\varepsilon$ (given $\gamma = \sigma^2/2$). Frames (a)-(f) represent the case $\sigma=7.5$. In frame (g) the dashed line represents the double power law with exponents $\alpha_1 = -0.2$ and $\alpha_2 = -1.5$. The kurtosis excess is defined as $\kappa = \mu_4 / \mu_2^2 - 3$, where $\mu_i$ is the $i$-th central moment of the cart velocity. 
}
\label{fig:sim_epsilon_dep}
\end{figure*}

The distribution of action points produced by the model decays exponentially, following the experimentally obtained distributions (Fig.~\ref{fig:sim_ap_distr}). This prompts that the suggested model captures the essence of the ``when-to-react'' mechanism employed by human subjects. The mismatch between the distributions around $\theta=0$ is apparently an artifact of the continuous approximation to the ``how-to-react'' mechanism. Specifically, due to the lack of highly precise corrections the system rarely reaches the close vicinity of the origin; the system trajectory thus leaves a noticeable gap around the origin (Figs.~\ref{fig:sim_phase_focus_gamma_low},~\ref{fig:sim_phase_focus_gamma_high}). 

Although the adopted optimal feedback approximation allows the model to capture well the peak velocity statistics observed in the experiments (Fig.~\ref{fig:sim_vmax_distr}), the analysis of the phase duration distributions confirms the need for a more advanced description of open-loop control (Fig.~\ref{fig:sim_ph_dur}). According to Eq.~\eqref{eq:act_ph_dur}, the duration of the active control phase of the model~\eqref{eq:model} is roughly constant for given $\sigma$, which is obviously unrealistic. As well, due to the lack of imprecise corrections the model demonstrates very few passive phases shorter than $\tau$, which also leads to increased number of passive phases longer than~$\tau$. A more adequate mathematical description of open-loop control presumably can eliminate this discrepancy.

However, even the rough approximation of the ``how-to-react'' mechanism allows the model to reproduce the experimental distributions of the stick angle and cart velocity regardless of the particular values of the parameter $\sigma$ (Fig.~\ref{fig:sim_distr_sigma}). The tails of both the $\theta$ and $\upsilon$ distributions generated by the model almost do not change for different $\sigma$. For high enough $\sigma$ the model reflects the bimodality of the stick angle distribution observed in the fast stick condition. The high peak of the velocity distribution at $\upsilon = 0$ is also captured for all tested $\sigma$. 

Finally, we touch on how the parameter $\varepsilon$ affects the system dynamics. The noise intensity $\varepsilon$ can be interpreted as the relative impact of the operator's aspiration to act compared to the resistance to change. Indeed, when the noise is absent, $\varepsilon=0$, the system cannot escape the vicinity of the $\theta$-axis. However, a non-zero noise intensity allows the system to eventually switch from the passive to the active phase. With increasing $\varepsilon$ the system spends less and less time in the passive phase. This point is illustrated in Fig.~\ref{fig:sim_epsilon_dep}. Given the noise intensity is small, the amplitude of the system fluctuations is extremely large (Fig.~\ref{fig:sim_phase_a}). Growing $\varepsilon$ leads to decreasing amplitude (Fig.~\ref{fig:sim_phase_b}), whereas the basic motion pattern remains unchanged. As long as $\varepsilon \ll 1$, the system trajectory remains smooth; $\varepsilon \sim 1$ (Fig.~\ref{fig:sim_phase_c}) marks the transition from the regular 
dynamics to the mostly random behavior (Fig.~\ref{fig:sim_phase_d}).

The match between the experimental and model distributions of $\theta$ and $\upsilon$ is stable with respect to variations of the noise intensity within a range of physically plausible values (Figs.~\ref{fig:sim_angle_distr},~\ref{fig:sim_velocity_distr}). Moreover, the height of the velocity distribution peak decreasing with $\epsilon$ may suggest that the subjects with low  $\%_{\textrm{drift}}$ (e.g., Subject~5) are characterized by relatively high values of $\varepsilon$. The effect of the noise intensity on the velocity distribution is further highlighted in Fig.~\ref{fig:sim_kurtosis}. The dependence of the velocity kurtosis excess on $\varepsilon$ is characterized by the double-power law decay, which persists for all tested values of $\sigma$. The power law exponent changes around $\varepsilon = 1$, suggesting two different modes of the system dynamics. First, for $\varepsilon \gtrsim 1$ the velocity kurtosis decays fast with $\varepsilon$, approaching the zero value (which indicates the Gaussian 
distribution). This mode corresponds to the essentially random motion, which prompts us to treat it as having little physical meaning. Second, for $\varepsilon \lesssim 1$ the velocity kurtosis remains high, which reflects the distribution peak at $\upsilon = 0$. This mode corresponds to intermittent control; noise here manifests itself primarily in the passive control phase, inducing the transition to the active phase. 

The results of theoretical and numerical analyses of the system~\eqref{eq:model} allow us to conclude that for a whole range of physically plausible parameter values the proposed model captures the control patterns exhibited by human subjects. 

\section{Discussion}
This paper illuminates that noise-driven control activation may be a core component of intermittent human control. We found that in overdamped stick balancing the subjects demonstrated clear intermittent control patterns. We hypothesize that human control behavior in the considered task is governed by two independent yet interacting mechanisms. The first, ``how-to-react'' mechanism is assumed to generate ballistic, open-loop corrections. The second, ``when-to-react'' mechanism operates during the passive control phase and intermittently activates the first one. The key idea of the paper is that control activation is not threshold-driven, but intrinsically stochastic, noise-driven. Specifically, we assume control triggering to result from the stochastic interplay between the operator's aspiration to keep the stick upwards and the resistance to interrupting the stick dynamics.

The model implementing the hypothesized mechanisms matches the key characteristics of human subjects' behavior. The phase trajectory exhibited by the model imitates the basic motion pattern of the overdamped stick under human control. Most importantly, the model closely reproduces the experimental distributions of the stick angle, cart velocity, and action points. This indicates that human subjects actually utilize noise-driven, not threshold-driven control activation mechanism. More subtle analysis suggests that a more advanced mathematical description of the open-loop system dynamics in the active phase should be developed in order to fully capture the intricate properties of the task dynamics. Overall, our results imply that noise-driven control activation plays a decisive role in human control at least in the considered task, and possibly in a wide class of human-controlled processes.

\subsection*{Overdamped stick balancing as a novel experimental paradigm}
This study is the first to experimentally investigate human control behavior in balancing a first-order unstable system. Previously the overdamped inverted pendulum and alike models have been used in studying the physics of human postural balance~\cite{milton2013intermittent}. Nevertheless, human control of the overdamped stick has never been investigated. Loram et al. examined human control of the virtual first-order load representing the massless inverted pendulum~\cite{loram2009visual}. However, such a load is inherently stable, which does not admit any direct implications for human control of unstable objects. 

The advantage of the experimental approach proposed here is that the intrinsic dynamics of the system under human control is ultimately simple, yet still unstable. The overdamped inverted pendulum has no dynamical properties that can be exploited in stabilizing the system, in contrast to the standard inverted pendulum~\cite{bottaro2008bounded, suzuki2012intermittent, asai2009model, asai2013learning}. More importantly, the human response delay supposedly does not contribute essentially to the dynamics of the control process. 

The processes traditionally studied in human motor control, such as underdamped stick balancing, have considerably more complex dynamics than the task at hand. On one hand, this may somehow limit the direct applicability of the findings reported here to such processes. On the other hand, the utmost simplicity of the present task enables one to identify and scrutinize potentially important control mechanisms whose presence may be obscured in the conventional experimental paradigms (e.g., due to sensorimotor noise, response delays, and complex intrinsic dynamics of a controlled system). As we demonstrate here, noise-driven control activation may be one of such previously overlooked mechanisms.

\subsection*{Noise-driven control activation: is there a threshold?}
The traditional threshold mechanism approximates a simple control algorithm: wait whenever the deviation is small, and act whenever the deviation is large. Threshold as a precise, fixed number is thus a somewhat artificial notion, so the modern literature on human control emphasizes that stochasticity of the threshold-based mechanism is necessary to capture human behavior. Hence, most available models of intermittent human control underline the crucial role of noise, either additive~\cite{bottaro2008bounded, milton2009discontinuous, gawthrop2011intermittent} or multiplicative~\cite{milton2009balancing, milton2013intermittent}. Still, even though noise can ``blur'' the threshold, resulting in some scatteredness of the action points, in such models control is de-facto triggered by the (noisy) controlled variable crossing the fixed threshold value.

The results of the present paper illuminate that control activation in humans may be not threshold-driven, but intrinsically stochastic, noise-driven. First, the experimentally found action point distribution reveals no distinct threshold value triggering human response (Fig.~\ref{fig:exp_ap_distr}). Exponential, not Gaussian decay of the action points suggests a highly stochastic, nonlinear control activation mechanism. Second, the distribution of action points observed in human subjects is reproduced by the model based on the assumption that control activation is a by-product of noise-mediated interaction between tendency to act and resistance to change (Fig.~\ref{fig:sim_ap_distr}). Third, the stick angle, cart velocity and peak cart velocity distributions are also well captured by the model, despite the approximate nature of the employed ``how-to-react'' mechanism. Furthermore, the match between the model and the experiments is observed for a range of the physically plausible parameter values, which confirms the robustness of the model. Overall, the found evidence for noise-driven activation in overdamped stick balancing raises a question whether similar mechanism is employed by humans in controlling more complex entities.

Previously it has been found that human subjects may exploit the stabilizing properties of multiplicative noise in order to handle the control of invered pendulum impeded by response delay~\cite{cabrera2002onoff, moss2003balancing}. However, the specific role of noise studied in Ref.~\cite{cabrera2002onoff} and related works is to disturb the feedback gain so that the closed-loop system intermittently switches between the stable and unstable dynamics. If the system is initially tuned to the unstable side of the stability boundary~\cite{milton2009discontinuous}, noise plays a constructive role, i.e., the system cannot be stabilized in the absence of noise. In regard to the latter point, the concept of noise-driven control activation proposed in this paper is similar to noise-induced stabilization studied by Milton, Cabrera et al. Still, whereas conventionally the noise component is introduced to mimic sensorimotor disturbances of small amplitude (e.g., due to limb tremor), we employ noise solely to mimic the stochasticity of the operator's decision process in the passive control phase. The similar interpretation of noise can be found, e.g., in the models of random switching between locally stable perceptions of ambiguous stimuli~\cite{bialek1995random,moreno2007noise}.

\subsection*{Implications and open directions}
We hypothesize that human control in overdamped stick balancing can be represented as repeated noise-driven triggering of the open-loop controller. However, the scope of the present paper is limited  mainly to the ``when-to-react'' mechanism, whereas the modeling framework for the ``how-to-react'' is still to be developed. The adopted optimal feedback approximation to open-loop control allows the model to capture the basic properties of the subjects' behavior. Still, a more adequate mathematical description of the active phase dynamics would presumably enable it to provide a deeper explanation of the experimentally observed dynamics. Particularly, we believe the noise-driven control activation, if coupled with stochastic open-loop mechanism, have the potential to explain anomalous dynamics of the systems controlled by humans, in particular, stick falls.

In regards to open-loop control, first, the experimental data should be studied in more detail to uncover the properties of the corrective movements generated by human subjects. Besides the already mentioned issue of highly imprecise and highly precise movements, the phase trajectories of the stick motion reveal that the subjects often interrupt the already launched correction, which results in multimodal fragments of the cart velocity profile. The properties of such fragments are to be analyzed using the variety of available methods~\cite{inoue2014wavelet}. Second, there is need for proper mathematical formalism capturing the stochasticity of the open-loop control mechanism. Even though the latter problem is indeed difficult to tackle, we feel that the overdamped stick balancing approach makes it simpler for one to address it compared to the standard experimental paradigms.

Another important aspect of human control left outside the scope of this work is learning. The experiments reported here were designed in such a way that the subjects' performance does not change considerably throughout the experiment trials. Nevertheless, in view of learning it appears noteworthy that the action point distributions exhibited by the most skilled and the least skilled participants are markedly different (Fig.~\ref{fig:exp_ap_distr}). This difference prompts that in learning to control the overdamped stick the subjects may adjust the parameter $\varepsilon$ in a search for some optimal value allowing for the accurate and at the same time energy-efficient control. The latter hypothesis requires separate consideration, which is also left for future studies.

The present results may have broader implications for the fields related to human control, e.g., the theory of car following. One may associate the process of keeping the stick upright with maintaining the comfortable headway to the car ahead by a car driver. Indeed, car following is a more complex process than stick balancing, yet some analogies can be drawn. The car following task is similar to stick balancing in that the process under human control is inherently unstable in the absence of operator actions. Similarly to stick balancing, human control in car following is also intermittent~\cite{wagner2003empirical}. In car following the action points in the headway \textemdash relative velocity phase plane are widely scattered~\cite{wagner2003empirical}, which can be linked to the action point variability in the present task (Fig.~\ref{fig:exp_ap_distr}). Finally, the Laplace distributions of the relative velocity obtained in car following~\cite{wagner2003empirical, wagner2012analyzing} are similar to the cart velocity distributions reported here. All these facts provide a preliminary basis for posing a hypothesis that noise-driven mechanism of recognizing the deviations from the ``optimal'' headway by the driver may be an essential factor underlying the fluctuations observed in car following.

According to our hypothesis, in balancing the overdamped stick the operator continuously observes the external process (i.e., the stick motion), and decides when and how exactly to interrupt it given the current circumstances. Similar processes (although in much more complex environments) are studied within the field of dynamic decision making, which focuses on the processes ``which require a series of decisions, where the decisions are not independent, where the state of the world changes, both autonomously and as a consequence of the decision maker's actions, and where the decisions have to be made in real time''~\cite{brehmer1992dynamic}. Similarly to overdamped stick balancing, in arguably any dynamic process involving human as a decision maker the procedure of detecting the deviations from the desired situation is stochastic in its nature. A system state either may be classified as acceptable with some probability,
 or may trigger the active behavior of a human observing the system. We thus believe that the concepts and models elaborated in the investigations of event-driven human control may potentially span across a general class of human-controlled processes.

\section*{Acknowledgments}
The authors thank Prof. Maxim Mozgovoy for invaluable help in conducting the preliminary experiments. The work was supported in part by the JSPS ``Grants-in-Aid for Scientific Research'' Program, Grant 24540410-0001.

\begin{appendices}
\renewcommand{\theequation}{\Alph{section}.\arabic{equation}}
\setcounter{equation}{0}
\renewcommand{\thefigure}{\Alph{section}.\arabic{figure}}
\setcounter{figure}{0}
\section{Motion equation of the overdamped inverted pendulum} \label{app:stick}
The mechanical system under consideration consists of the movable cart and the stick of length $l$ (Fig.~\ref{fig:stick}). Without loss of generality we assume that the stick mass $m$ is concentrated at its upper end. The bottom end of the stick and the cart are connected via the frictionless pivot. The system is assumed to be embedded in a viscous environment characterized by the coefficient of viscous friction $k$.

In the non-inertial reference frame attached to the cart the dynamics of the system are described by the equation
\begin{equation}
  ml^2\ddot{\theta} = mgl\sin\theta - ml\dot{\upsilon}\cos\theta - k(l\upsilon\cos\theta + l^2 \dot{\theta}),
\label{eq:app1_moments}
\end{equation}

We divide both sides of Eq.~\eqref{eq:app1_moments} by constant factor $mgl$ and then rescale time $t$ and cart velocity $\upsilon$
\begin{equation*}
 t \to t \tau,\quad \upsilon \to \upsilon \frac{l}{\tau}, \quad \textrm{where } \tau = \frac{kl}{mg},
\end{equation*}
so that Eq.~\eqref{eq:app1_moments} reads
\begin{equation}\label{eq:app1_nondim}
  \frac{m^2 g}{k^2 l}\ddot{\theta} = \sin\theta - \frac{m^2 g}{k^2 l}\dot{\upsilon}\cos\theta - (\upsilon\cos\theta + \dot{\theta}),
\end{equation}

Given that the cart motion occurs on the spatial scale of the stick length $l$ and the same time scale as the stick angular motion, the terms of Eq.~\eqref{eq:app1_nondim} containing $\ddot \theta$ and $\dot \upsilon$ contribute little to the system dynamics in the limit of high viscosity ($k \gg m \sqrt{\frac{g}{l}}$) and thus can be neglected. Returning to the original variables, Eq.~\eqref{eq:app1_nondim} finally reads
\begin{equation}\label{eq:app1_final}
  \tau \dot{\theta} = \sin\theta - \frac{\tau}{l}\upsilon\cos\theta.
\end{equation}

\setcounter{equation}{0}
\setcounter{figure}{0}
\section{Optimal feedback approximation to open-loop control}  \label{app:openloop}
In this appendix we derive the continuous approximation for the open-loop actions of human operator in controlling the overdamped stick
\begin{equation}
 \tau \dot \theta = \theta - \frac{\tau}{l} \upsilon.
\label{eq:app2_stick}
\end{equation}

We employ the function
\begin{equation*}
  F(\upsilon,\dot{\upsilon},\theta) = \frac{\tau^2}{2l^2}\left(\upsilon^2 + \tau_m^2\dot{\upsilon}^2\right) + \frac{\theta^2}{2\theta_m^2}
\end{equation*}
to measure the current state of the system in its motion near the equilibrium $\upsilon=0,\,\theta = 0$. The parameter $\theta_m$ denotes the characteristic stick angle regarded by the operator as large enough to correct the stick position. The time scale $\tau_m$ can be interpreted as the characteristic duration of a single corrective movement.

A possible course of the future operator actions $v(t'),\,t'\geq t$ aimed at returning the system~\eqref{eq:app2_stick} from the current state $\theta(t) = \theta_0$, $\upsilon(t)=0$ to the desired state $\theta=0$, $\upsilon=0$ can be then characterized by the integral measure
\begin{equation}\label{eq:app2_loss}
 \mathcal{F}\{\upsilon\} = \int\limits_t^\infty\left[\frac{\tau^2}{2l^2}\left(\upsilon^2 + \tau_m^2\dot{\upsilon}^2\right) + \frac{\theta^2}{2\theta_m^2}\right]dt'
\end{equation}
where for a given $\upsilon(t')$ the time dependence $\theta(t')$ of the stick angle is determined by Eq.~\eqref{eq:app2_stick}. Integral~\eqref{eq:app2_loss} quantifies the priority of possible operator actions. Assuming the operator to be able to perfectly predict and measure the system dynamics, the optimal strategy $\upsilon_\text{opt}$ is the solution of the optimization problem 
\begin{equation}\label{eq:app2_opt}
\upsilon_\text{opt} = \argmin_{\upsilon\{t'\}}\mathcal{F}\{\upsilon\}
\end{equation}
subject to the system dynamics equation~\eqref{eq:app2_stick}, the initial and terminal conditions
\begin{equation*}
\upsilon(t) = 0\,,\quad \theta(t) = \theta_0\,,\quad \upsilon(\infty) = 0\,,\quad \theta(\infty) = 0\,.
\end{equation*}

We reduce the problem~\eqref{eq:app2_opt} to a standard variational problem using the technique of Lagrange multipliers,
\begin{equation*}
\label{eq:app2_Lag_problem}
\upsilon_\text{opt} = \argmin_{\upsilon\{t'\},\theta\{t'\},\mu\{t'\}}\ \int\limits_t^\infty F_L(\upsilon,\dot{\upsilon},\theta,\dot{\theta},\mu)\,dt',
\end{equation*}
where
\begin{equation*}
\label{eq:app2_Lag_function}
 F_L(\upsilon,\dot{\upsilon},\theta,\dot{\theta},\mu) = F(\upsilon,\dot{\upsilon},\theta) + 
\mu \left[\tau\dot{\theta} - \theta + \frac{\tau}{l} \upsilon\right].
\end{equation*}
The Lagrange equation 
\begin{equation*}
\frac{\partial F_L}{\partial q} - \frac{d}{dt'}\frac{\partial F_L}{\partial \dot{q}} = 0 \quad\text{for $q=\upsilon$, $\theta$, $\mu$}
\end{equation*}
yields the equations determining the optimal actions of the operator
\begin{equation}
\label{eq:app2_Lag_solution}
\begin{aligned}
	\tau_m^2 \ddot{\upsilon} & = \upsilon + \frac{l}{\tau}\mu\,, \\
	\tau\dot{\mu} & = \frac{1}{\theta_m^2}\,\theta -\mu \,,\\
	\tau\dot{\theta} & = \theta - \frac{\tau}{l}\upsilon\,. 
\end{aligned}  
\end{equation}
The eigenvalues of the matrix corresponding to the linear system~\eqref{eq:app2_Lag_solution} are the solutions of the equation
\begin{equation}\label{eq:app2_eig_eq}
(\tau_m^2 \lambda^2 -1)(\tau^2 \lambda^2 -1) = -\frac{1}{\theta_m^{2}}\,.
\end{equation}
Equation~\eqref{eq:app2_eig_eq} has four roots subject to the condition
\begin{multline}\label{eq:app2_eig_general}
\lambda^2 = \frac1{2\tau_m^2\tau^2} \Bigg\{ (\tau_m^2 + \tau^2) \\
	\pm i\left[\frac{4\tau_m^2\tau^2}{\theta_m^{2}}-(\tau_m^2 - \tau^2)^2\right]^{1/2} \Bigg\}.
\end{multline}
Appealing to the experimental results and physical meaning of the parameters $\tau_m$ and $\theta_m$, we make the estimates
\begin{equation}\label{eq:app2_assumptions}
\tau_m\sim\tau \quad\text{and}\quad\theta_m\ll 1\,,
\end{equation} 
which enable us to simplify Exp.~\eqref{eq:app2_eig_general},
\begin{equation}\label{eq:app2_eig_simplified}
\lambda^2 = \pm \frac{i}{\tau\tau_m \theta_m} \,.
\end{equation}
Equation~\eqref{eq:app2_eig_simplified} possesses the roots
\begin{subequations}
\label{eq:app2_lambda}
\begin{align}
\label{eq:app2_lambda_neg}
	\lambda_{1,2} = \frac{1}{(2\theta_m\tau_m\tau)^{1/2}}(-1 \pm i), \\
\label{eq:app2_lambda_pos}
	\lambda_{3,4} = \frac{1}{(2\theta_m\tau_m\tau)^{1/2}}(1 \pm i).
\end{align}  
\end{subequations}

The minimization problem~\eqref{eq:app2_opt} is a temporal boundary value problem: the solution of the problem~\eqref{eq:app2_opt} is determined by the initial and the target system position. Within the accepted model the terminal conditions $\upsilon(\infty) = 0$, $\theta(\infty) = 0$ enable us to disregard the eigenvectors matching the eigenvalues $\lambda_{3,4}$ due to $\text{Re\,}\lambda_{3,4} >0$. This reduces the original boundary value problem to an initial value problem. The solution then can be constructed using only the current system state, $v(t) = 0,\,\theta(t) = \theta_0$. Therefore, a dynamical system possessing the eigenvalues $\lambda_{1,2}$ specified by Exp.~\eqref{eq:app2_lambda_neg} can equivalently describe the dynamics of the system~\eqref{eq:app2_stick} under control of human operator aiming to compensate for a detected stick deviation while minimizing the loss function~\eqref{eq:app2_loss}. Specifically, the system
\begin{subequations}
\label{eq:app2_solution}
\begin{align}
\label{eq:app2_solution_a}
\dot{\upsilon} & = \alpha l \theta - \beta \upsilon\,,
\\
\label{eq:app2_solution_b}
\tau \dot{\theta} & = \theta - \frac{\tau}{l}\upsilon\,,
\end{align}
\end{subequations}
has the eigenvalues~\eqref{eq:app2_lambda_neg} given that $\beta = \sqrt{2} (\theta_m\tau_m\tau)^{-1/2}$ and $\alpha = \beta^2 / 2$. 

The optimal feedback defined by Eq.~\eqref{eq:app2_solution_a} can be treated as an approximation to open-loop control. The conventional understanding of the latter implies that, once launched, the corrective movement is not interrupted until fully executed. Similarly to open-loop control in this traditional sense, the operator acting as described above calculates the response only once and then does not change the established control pattern. Indeed, assume the operator ``solves'' an optimization problem to generate the control movement each time some stick deviation $\theta_0$ triggers the active response. Then, according to Bellman's principle of optimality, any potential corrections of the calculated response during its execution cannot improve the overall quality of that response. Therefore such corrections are not implemented by the optimally acting operator. Of course, it is a very strong assumption that the operator generates an open-loop control response $\upsilon_\text{opt}(t')$ for $t'>t$ exactly 
in a way that it produces the trajectory exhibited by the system~\eqref{eq:app2_solution}. That is why we would like to underline that the proposed control mechanism is just an approximation to the experimentally observed behavior, and that the appropriate mathematical formalism should be developed to capture the open-loop nature of human control.

\end{appendices}

\bibliography{library}

\end{document}